\documentclass[12pt,twoside]{article}
\usepackage{amsmath,amssymb,amsfonts,euscript}

\newtheorem{theorem}{Theorem}
\newtheorem{lemma}{Lemma}
\newtheorem{proposition}{Proposition}
\newtheorem{corollary}{Corollary}
\newtheorem{definition}{Definition}
\newtheorem{remark}{Remark}

\newcommand{\qt}[2]{\text{\raisebox{6pt}{#1\!\!\raisebox{-3mm}{\Huge/}
             \!\raisebox{-3mm}{\normalsize#2}}}}
\newcommand{\rst}[2]{#1\raisebox{-1mm}{{}\!\Large$|$}\raisebox{-2mm}{$#2$}}
\newcommand{\ud}{\mathrm{d}}
\newcommand{\mymap}{\raisebox{19572sp}{$\shortmid$}
              \!\!\!\!\:-\!\!\!-\!\!\!-\!\!\!-\!\!\!-\!\!\!-\!\!\!-\!\!\!-
              \!\!\!-\!\!\!-\!\!\!\longrightarrow}
\newcommand{\crl}{[\:\!\![}
\newcommand{\crr}{]\:\!\!]}
\newcommand{\bcrl}{\bigl[\;\!\!\!\bigl[}
\newcommand{\bcrr}{\bigr]\;\!\!\!\bigr]}
\newcommand{\ys}{{\displaystyle \rlap{\raisebox{1pt}{{\normalsize /}}}
             \raisebox{160000sp}{$y$}}}
\newcommand{\ds}{\rlap{\raisebox{1pt}{/}}\;\!\!\partial}
\newcommand{\gm}{\displaystyle {\gamma}}
\newcommand{\om}{\displaystyle {\omega}}
\newcommand{\yst}[1]{\ys^{\text{\raisebox{-2pt}{$(#1)$}}}}
\newcommand{\dst}[1]{\ds^{(#1)}}
\newcommand{\Di}{{\mathrm {Di}}}
\newcommand{\Rac}{{\mathrm {Rac}}}
\begin{document}

%
%                   Some  properties of massless particles
%                         in arbitrary dimensions
%
\title{\bf Some properties of massless particles \\
in arbitrary dimensions\thanks{To be published
in {\it Reviews in Mathematical Physics}}
           \vspace{1cm} }
\author{\sc{Mourad} LAOUES  \bigskip \\
 {\footnotesize Laboratoire Gevrey de Math\'ematique Physique} \\
{\footnotesize UNIVERSIT\'E DE BOURGOGNE} \\
{\footnotesize 9, avenue Alain Savary }\\
{\footnotesize B.P. 400, F-21011 Dijon Cedex, France} \\
\texttt{\footnotesize laoues@u-bourgogne.fr, physmath@u-bourgogne.fr}}
\date{\footnotesize {January 26, 1998}}
\maketitle

%
%%%%%%%%%%%%%%%%%%%%%%%%%          Abstract          %%%%%%%%%%%%%%%%%%%%%%%%%%
%
\begin{abstract}
Various properties of two kinds of massless representations of the
$n$-con\-formal (or $(n+1)$-De Sitter) group $\tilde{G}_n=\widetilde{SO}_0(2,n)$
are investigated for $n\ge2$. It is found that, for space-time dimensions
$n\ge3$, the situation is quite similar to the one of the $n=4$ case
for $S_n$-massless representations of
the $n$-De~Sitter
group $\widetilde{SO}_0(2,n-1)$. These representations are the restrictions
 of the singletons of $\tilde{G}_n$.
The main difference is that they are not contained in the tensor product of
two UIRs with the same sign of energy when $n>4$, whereas it is the case
for another kind of massless representation. Finally some examples of
Gupta-Bleuler triplets are given for arbitrary spin and $n\ge3$.
\end{abstract}
\thispagestyle{empty}
\newpage

\begin{center}
\section{Introduction} \label{S:intro}
\end{center}

  The (ladder) representations $D(s+1,s,\epsilon s), 2s\in\mathbb{N}$ and
$|\epsilon|=1$ of the universal covering $\tilde{C}_4 = \widetilde{SO}_0(2,4)$
of the conformal group remain irreducible when restricted to the universal
covering $\tilde{P}_4 = \widetilde{SO}_0(1,3)\ltimes T_4$ of the Poincar\'e
group and each non trivial positive energy representation of the conformal group
with that property is equivalent to one of them. However the restriction to
the universal covering $\tilde{S}_4$ of the De Sitter group is irreducible only
if $s>0$; indeed one has:
$$
  \rst{D(s+1,s,\epsilon s)}{\tilde{S}_4}=
     \begin{cases}
     D(s+1,s) &\text{\qquad if $s>0$}; \\
     D(1,0)\oplus D(2,0) &\text{\qquad if $s=0$}.
     \end{cases}
$$
These representations are called massless (relatively to the De Sitter group)
for a variety of reasons \cite{AFFS81}.
In the present paper we call them $S_4$-massless representations of the
De Sitter group $\tilde{S}_4 = \widetilde{SO}_0(2,3)$ because, as indicated in
\cite{AFFS81,FF83,FFG86} they satisfy the following
\emph{masslessness conditions}:\\
\indent (a) They contract smoothly to a massles discrete helicity 
representation of the
Poincar\'e group $\tilde{P}_4 = \widetilde{SO}_0(1,3)\ltimes\mathbb{R}^4$;\\
\indent (b) Any massless discrete helicity representation $U^P$ of the
 Poincar\'e group
has a unique extension to a UIR $\Hat{U}$ (called $C_4$-massless representation
in this paper) of the conformal group $\tilde{C}_4 = \widetilde{SO}_0(2,4)$.
The restriction of $\Hat{U}$ to the De Sitter group is precisely one of the
massless representations of $\tilde{S}_4$ recalled above;\\
\indent (c) For spin $s\ge1$ one may construct a gauge theory on the 
Anti-de Sitter
space for massles particles, quantizable only by the use of an indefinite
metric and a Gupta Bleuler triplet;\\
\indent (d) The massless representations in question distinguish themselves 
by the fact
that the physical signals propagate on the Anti-de Sitter light cone.

  Other interesting representations of $\tilde{S}_4$ are the Dirac singletons
$\Di= D(1,\frac{1}{2})$ and $\Rac= D(\frac{1}{2},0)$ (which are also
$C_3$-massless representations in the sense defined below).
Some of their properties are:
\begin{enumerate}
  \item  Dirac singletons are, up to equivalence, the only unitary irreducible
positive energy representations of $\tilde{S}_4$ which remain irreducible when
restricted to the universal covering $\tilde{L}_4$ of the Lorentz group;
  \item  In the limit of zero curvature (of the De Sitter space
$\tilde{S}_4/\tilde{L}_4$) they contract to unitary irreducible representations
(UIR) of $\tilde{P}_4$ that are trivial on the translation part $T_4$;
  \item  Let $\chi(\mu_1)\otimes\pi(\mu_2)$ denote the IR (up to equivalence),
with highest weight $(\mu_1,\mu_2)$ of the universal covering $\tilde{K}_4$
of the maximal compact subgroup of the De Sitter group. Then the restriction to
$\tilde{K}_4$ of the Dirac singletons UIRs of the De Sitter group is given by
$$ \rst{D(\frac{1}{2}+s,s)}{\tilde{K}_4}=
\bigoplus_{l\in\mathbb{N}}\chi(-[\frac{1}{2}+s+l])\otimes\pi(s+l),
 \;s=0\text{\;or\;}\frac{1}{2}. $$
  \item  Finally the Dirac singletons satisfy the following \cite{FF78}:
 \begin{align}
 \Rac\otimes \Rac=& \,\;\bigoplus_{s\in\mathbb{N}}D(s+1,s);   \notag \\
 \Rac\otimes \Di\; =& \bigoplus_{s-\frac{1}{2}\in\mathbb{N}}D(s+1,s); \notag \\
 \Di \otimes \Di\; =& \bigoplus_{s-1\in\mathbb{N}}D(s+1,s)\oplus D(2,0). \notag
 \end{align}
\end{enumerate}
 Note \cite{AFFS81} that the Dirac singletons are not massless representations
of the De Sitter group. But if one considers $S_4$ as the conformal group of
the 3-dimensional Minkowski space then the Dirac singletons are massless, i.e.
their restriction to the corresponding Poincar\'e group $P_3$ is irreducible
\cite{AFFS81,AL97,FFG86}. In this case it is clear from the context what kind
of masslessness is considered. However, for general $n$, some confusion
may arise. To avoid it we shall introduce a prefix to the word ``massless"
(see definition \ref{D: CS-mass}), to distinguish between ``conformal
masslessness" and ``De Sitter masslessness" in any dimension, to precise which
group we are representing.

  A common property to both types of massless representations is the existence
of Gupta-Bleuler (GB) quantization; see for example
\cite{AFFS81,BFrH(S)83,Fr85,H84}.

  The purpose of this work is to continue the study performed in \cite{AL97}
and more specifically to look for properties of maslessness (both types) which
persist when the space-time dimension becomes an arbitrary integer $n\ge2$.
In Section \ref{S:gen} we fix the notations and recall some results.
In Section \ref{S: Irr-Ctr} we discuss the irreducibility of a massless
representation of the $n$-conformal group when restricted to the
$(n+1)$-Lorentz group and its contractibility to UIRs of the $n$-Poincar\'e
group. Reduction to the maximal compact subgroup of the conformal group is
studied in Section \ref{S: Red on K}. Finally Dirac singletons and
Gupta-Bleuler triplets are treated in (respectively) Sections \ref{S: DS}
and \ref{S: GB}.
It is found that almost all the properties of massless representations in
dimension $n=4$ are conserved when $n\ge3$; however the property that massless
representations are, when $n=4$, contained in the tensor product of two
positive energy UIRs (of the De Sitter group) fails for general $n$.

After a first version of this paper was written appeared a preprint
\cite{FerFr98b} with somewhat different conclusions, based on a less-demanding
notion of masslessness in higher dimensions. Since we need the definitions
and results of this paper to compare both notions, we shall discuss this
point at the end of the paper.

%
%                              2  Generalities
%
\begin{center}
\section{Generalities} \label{S:gen}
\end{center}

   We suppose $ n \ge 2 $.
Let $ \mathbb{R}^{1,n-1} $ be the $n$-dimensional Minkowski space-time,
$T_n$ its group of translations, $ L_n = SO_0(1,n-1) $ the $n$-Lorentz group,
$ P_n = L_n \ltimes T_n $ the $n$-Poincar\'e  group and $ S_n = SO_0(2,n-1) $
the $n$-De Sitter group. We write $ \mathcal{T}_{n} $, $ \mathcal{L}_{n} $,
$ \mathcal{P}_{n} $ and $ \mathcal{S}_n $ the corresponding Lie algebras.

Let $G_n = SO_0(2,n)$. The preceding groups may be considered as subgroups
of $G_n$. Indeed let $\bigl(M_{ab}\bigr)_{-1\le a<b\le n}$ be a basis of the
Lie algebra $\mathcal{G}_n$ of $G_n$ such that:
  \begin{align}
   M_{ab} &= -M_{ba}                  \\
\mbox{and \qquad } [M_{ab},M_{cd}] &=
\eta_{ad}M_{bc}+\eta_{bc}M_{ad}-\eta_{ac}M_{bd}-\eta_{bd}M_{ac} \\
\mbox{where  \quad\qquad\qquad}     \eta &=
              \left(\!\! \begin{array}{c|c}
                        \mathbf{\;\;1}_2\! &             \\
                                        \hline
                                           & \!\!-\mathbf{1}_n \\
              \end{array}  \!\!\right)     \notag
  \end{align}

\noindent We now imbed the above mentioned Lie algebras in $\mathcal{G}_n$ in
the following way:

     \begin{align}
 \mathcal{T}_n &= \langle \;  M_{-1,\alpha} + M_{\alpha,n} ,
\quad 0 \le \alpha \le n-1 \; \rangle \notag \\
 \mathcal{L}_n &= \langle \; M_{\alpha \beta} ,\qquad 0 \le \alpha ,
\beta \le n-1 \; \rangle  \notag   \\
 \mathcal{S}_n &= \langle \; M_{ab} ,\qquad -1 \le a,b \le n-1 \; \rangle .
 \notag     \\
 \mbox{Let: \quad\qquad\qquad}
 \overline{\mathcal{T}}_n &= \langle \; M_{-1,\alpha} - M_{\alpha,n} ,
\quad 0 \le \alpha \le n-1 \; \rangle  \notag \\
 \mathcal{D}_n &= \langle \; M_{-1,n} \; \rangle \notag
     \end{align}

\noindent and $\overline{T}_n$ (resp. $D_n$) the connected subgroup of $G_n$,
the Lie algebra of which is $\overline{\mathcal{T}}_n$ (resp.$\mathcal{D}_n$).
Then we define the $n$-conformal group of $\mathbb{R}^{1,n-1}$ as the closed
subgroup $C_n$ of $G_n$ generated by $T_n$, $L_n$, $D_n$
and $ \overline{T}_n$. $C_n$, $G_n$ and $S_{n+1}$ are locally isomorphic;
one has, if $\mathcal{C}_n$ denotes the Lie algebra of $C_n$:
 $$ \mathcal{C}_n = \mathcal{G}_n = \mathcal{S}_{n+1}. $$
Note that with our definition by ``conformal group of $\mathbb{R}^{1,1}$"
we mean here the group $SO_0(2,2)/\mathbb{Z}_2$.

Let $G$ a Lie group. We denote by $\tilde{G}$ be the spinorial covering of $G$
when $G$ is isomorphic to $L_3$ or $P_3$ and the universal covering otherwise.
Let $U$ be a non trivial highest weight unitary representation of
$\tilde{C}_n$.
%
%  Definition 1
%
  \begin{definition} \label{D: CS-mass}
  We say that $U$ is \emph{$C_n$-massless}\footnote{massless relatively to the
$n$-conformal group.} whenever $U$ and $\rst{U}{\tilde{P}_n}$ are irreducible.
  We say that $U$ is \emph{$S_{n+1}$-massless}\footnote{massless relatively
to the $(n+1)$-De Sitter group.} whenever $U$ is a restriction to
$\tilde{S}_{n+1}\simeq\tilde{C}_n$ of a $C_{n+1}$-massless representation of
$\tilde{C}_{n+1}$.
  \end{definition}
Note that the notions of $C_n$-massless and $S_n$-massless refer to
$n$-dimensional space-times $\mathbb{R}^{1,n-1}$ ($n$-Minkowski space) and
$\tilde{S}_n/\tilde{L}_n$ (which we call the $n$-De Sitter space, though
$n$-Anti De Sitter space might be a more appropriate expression).
The conformal group of both of them is $\tilde{C}_n$, locally isomorphic
to $\widetilde{SO}_0(2,n)$  while the invariance groups are respectively
$P_n$ and $\tilde{S}_n=\widetilde{SO}_n(2,n-1)$. For example usual massless
particles in 4-Minkowski space, for the Poincar\'e group, are in fact
$C_4$-massless (under extension to the conformal group
$SO_0(2,4)/\mathbb{Z}_2$) and $S_4$-massless (under deformation to the
De Sitter group $SO_0(2,3)$); usual Dirac singletons are $C_3$-massless,
and their restrictions to $SO_0(2,2)$ (reducible in a sum of two) are
$S_3$-massless.

  For simplicity we identify the group representation $U$, the Lie algebra
representation it defines $\ud U$ and the extension of the latter to
$\mathcal{U}(\mathcal{C}_n)$ and denote
\( I = \ker_{\mathcal{U}(\mathcal{C}_n)}(U) \). Then one has \cite{AL97}:

%
%  Theorem 1
%
 \begin{theorem} \label{T: FR}
 \begin{equation} \label{E: FR}
\!U\;is\;C_n\mbox{-}massless \! \Longleftrightarrow \! \eta^{cd}M_{ac}M_{bd}-
  \frac{n}{2}M_{ab}+\frac{2}{n+2}\eta_{ab}\mathbf{C}_2 = 0 \pmod{I}
 \end{equation}
$$\forall a,b \in \{-1,\ldots,n\}$$
where $\mathbf{C}_2$ is the Casimir operator and where we have used
the Einstein summation con\-vention\footnote{summation on repeated indices.}.
  \end{theorem}

%
%  Definition 2
%
 \begin{definition} \label{D: FR}
We call the right hand side of the preceding equivalence \eqref{E: FR}
the \emph{fundamental relation (FR)}.
  \end{definition}

%
%        3  Irreducibility of $U$ when restricted to $\tilde{L}_{n+1}$
%             and its contractibility to UIRs of $\tilde{P}_{n+1}$
%
\begin{center}
\section{Irreducibility of $U$ when restricted to $\tilde{L}_{n+1}$
and its contractibility to UIRs of $\tilde{P}_{n+1}$} \label{S: Irr-Ctr}
\end{center}

%
%    3.1  Irreducibility of $U$ when restricted to $\tilde{L}_{n+1}$
%
\subsection{Irreducibility of $U$ when restricted to $\tilde{L}_{n+1}$ }
\label{Sb: Irr}

  The following proposition is a characterization of UIRs $U$ which remain
irreducible when restricted to $\tilde{L}_{n+1}$.

%
%  Proposition 1
%
\begin{proposition} \label{P: Irr}
 Let $(U,\mathcal{H})$ be a highest weight UIR of $\tilde{C}_n$. Then:
  \begin{equation} \label{E: Irr}
 \rst{U}{\tilde{L}_{n+1}} \mbox{is irreducible} \quad \Longleftrightarrow
 \quad U \;   \mbox{satisfies the FR}.
  \end{equation}
\end{proposition}

\emph{Proof. } Assume that the restriction  $\rst{U}{\tilde{L}_{n+1}}$ is
irreducible. Then both of the Casimir operators of $\mathcal{G}_n$ and
$\mathcal{L}_{n+1}$ are sent to the scalars by $U$ and
$\rst{U}{\tilde{L}_{n+1}}$ respectively. It follows that the difference of
these operators is also sent to the scalars and thanks to the adjoint action of
$\mathcal{G}_n$ one obtains the FR. The converse is proved in \cite{AL97}.

  It easily follows:
%
%  Corollary 1
%
\begin{corollary} \label{C: Irr}
  If $U$ is $C_n$-massless then $\rst{U}{\tilde{L}_{n+1}}$ is irreducible.
\end{corollary}

%
%    3.2  A contraction of $C_n$-massless representations
%
\subsection{A contraction of $C_n$-massless representations} \label{Sb: Ctr}

  Consider a family $(\mathbf{S}_{\rho})_{0<\rho\le1}$ of operators defined
on the underlying vector space $V_n$ of $\mathcal{S}_{n+1}=\mathcal{G}_n$ by:
      \begin{align}
 \mathbf{S}_{\rho}(M_{\alpha\beta})
&= M_{\alpha\beta}  &&\text{if}\qquad 0 \le \alpha,\beta \le n  \\
 \mathbf{S}_{\rho}(M_{-1\alpha})
&= \sqrt{\rho}\:M_{-1\alpha} &&\text{if}\qquad 0 \le \alpha \le n.  \notag
      \end{align}
It defines a contraction of $\mathcal{S}_{n+1}$ to $\mathcal{P}_{n+1}$.
We are using here the notion of contractions of representations
(on Hilbert spaces) given in \cite{AFFS81} (see also \cite{AL97}).

Let $\mathcal{G}_n^\rho$ be the Lie algebra isomorphic to $\mathcal{G}_n$
defined by the bracket:
     \begin{equation}
 [x,y]_\rho = \mathbf{S}_\rho^{-1}[\mathbf{S}_{\rho}x,\mathbf{S}_{\rho}y],
\qquad x,y \in V_n
     \end{equation}
and let $U_\rho$ the representation of $\mathcal{G}_n^\rho$ defined on the
corresponding space $\mathcal{H}$ by:
     \begin{equation}
 U_\rho(x) = \mathbf{Z}_\rho^{-1} \circ U(\mathbf{S}_{\rho}x)
\circ \mathbf{Z}_\rho,\quad x \in V_n,
   \end{equation}
where \((\mathbf{Z}_\rho)_{0<\rho\le1}\) is a continuous family of closed
invertible operators of $\mathcal{H}$, $\mathbf{Z}_1$ being the identity. We
choose them here such that:
     \begin{equation}
\mathbf{Z}_\rho^{-1}U(M_{\alpha\beta})\mathbf{Z}_\rho =
U(M_{\alpha\beta}),\quad 0 \le \alpha,\beta \le n.
     \end{equation}
Thus one has:
     \begin{equation}
U_\rho(M_{\alpha\beta}) = U(M_{\alpha\beta}), \qquad 0 \le \alpha,\beta \le n
     \end{equation}
and, using the FR:
     \begin{align}
U_\rho(M_{-1\alpha})U_\rho(M_{-10})
&= \mathbf{Z}_\rho^{-1}U(\mathbf{S}_{\rho}M_{-1\alpha})
  U(\mathbf{S}_{\rho}M_{-10})\mathbf{Z}_\rho    \\
&= \rho\mathbf{Z}_\rho^{-1}U(M_{-1\alpha}M_{-10})\mathbf{Z}_\rho  \notag  \\
&= \rho U(\sum_{k=1}^{n}M_{\alpha k}M_{0 k}-\frac{n}{2}M_{0\alpha}-
   \frac{2}{n+2}\eta_{0\alpha}\mathbf{C}_2).  \notag
     \end{align}
Thus the operator $U_\rho(M_{-10}^2)$ has limit zero when $\rho \to 0$, in the
sense that it sends a dense subspace to \{0\} when $\rho \to 0$. It follows
that the limit of $U_\rho(M_{-1\alpha})$,$0 \le \alpha \le n$, is zero too.

  If one chooses $\rho$ to be the curvature of the space $S_{n+1}/L_{n+1}$
then one can write, from what precedes:

%
%  Proposition 2
%
\begin{proposition} \label{P: Ctr}
  In the limit of zero curvature the contracted \mbox{$C_n$-massless}
representation is trivial on $T_{n+1}$, the translation part of
$\tilde{P}_{n+1}$.
\end{proposition}

%
%     4  Reduction of U on the maximal compact subgroup of $\tilde{C}_n$
%
\begin{center}
\section{Reduction of $U$ on the maximal compact subgroup of
$\tilde{C}_n$} \label{S: Red on K}
\end{center}

The following results are proved in \cite{AL97}:

%
%  Theorem 2
%
\begin{theorem} \label{T: Caract of mass 0}
  Let $ \lambda = (\lambda_1,\ldots,\lambda_r)$, $r=[\![\frac{n+2}{2}]\!]$
(the integer part of $\frac{n+2}{2}$), the highest weight (HW) of the
\mbox{$C_n$-massless} representation $U$.
 Then there exists a real number $s$ such that:

  \begin{equation}
 \lambda_1 = -s-\frac{n-2}{2}\; \text{and}\; \lambda_2 =
\cdots = \lambda_{r-1} = |\lambda_r| = s
  \end{equation}
\[
\text{where:\qquad}
        \begin{cases}
  s > 0            &\text{if $n=2$;} \\
  2s\in\mathbb{N}  &\text{if $n$ is even and $n\ge4$;} \\
  \lambda_r\ge0\; \text{and}\; s=0\; \text{or}\; 1/2 &\text{if $n$ is odd.}
        \end{cases}
\]

\end{theorem}

%
%  Proposition 3
%
\begin{proposition} \label{P: Red K}
  Let $\mathfrak{k}_n = \mathfrak{so}(2) \oplus \mathfrak{so}(n)$ the maximal
compact subalgebra of $\mathcal{C}_n$ and let
$\chi(\mu_1)\otimes\pi(\mu_2,\ldots,\mu_r)$ be an IR of $\mathfrak{k}_n$
with HW $\mu=(\mu_1,\ldots,\mu_r)$. Then one has:
    \begin{equation}
 \rst{U}{\mathfrak{k}_n} = \bigoplus_{l=o}^{\infty}
\chi(-s-\frac{n-2}{2}-l)\otimes\pi(s+l,s,\ldots,s,\epsilon s),
     \end{equation}
where $|\epsilon|=1$ (resp. $\epsilon=1$) if $n$ is even (resp. odd).
\end{proposition}

  Thus $C_n$-massless representations are very degenerate and are,
in some sense, ``singleton'' representations.

%
%                          5   Dirac singletons
%
\begin{center}
\section{Dirac singletons} \label{S: DS}
\end{center}

%
%  Definition 3
%
\begin{definition}  \label{D: DS}
  A \emph{positive} (resp. \emph{negative}) \emph{energy} representation of
$\tilde{S}_n$ is a lowest (resp. highest) weight representation.
  We say that $U$ and $U'$ are \emph{Dirac singletons (DS)} if a
$S_n$-massless representation of $\tilde{S}_n$ occurs in the reduction of the
product $U \otimes U'$ and if $U$ and $U'$ have the same sign of energy.
\end{definition}

  It has been proved by M. Flato and C. Fr\o nsdal in \cite{FF78} for the $n=4$
case that the (irreducible and unitary) representations $\Di=D(1,1/2)$ and
$\Rac=D(1/2,0)$ are Dirac singletons and that the product
$(\Di \oplus \Rac) \otimes (\Di \oplus \Rac)$ reduces to a direct sum of
$S_4$-massless representations of $\widetilde{SO}_0(2,3)$. It is interesting
to note that $\Di$ and $\Rac$ are $C_3$-massless representations of
$\widetilde{SO}_0(2,3)$. Unfortunately when $n\ge5$ things behave differently;
the next proposition treats this case.

%
%  Proposition 4
%
\begin{proposition} \label{P: S for n>4}
  Assume $n\ge5$.
  Let $U$ and $U'$ be DS. Then only a finite number of $S_n$-massless
representations of $\tilde{S}_n$ can occur in the reduction of the product
$U \otimes U'$. Moreover $U$ and $U'$ can not be simultaneously unitary.
\end{proposition}

\emph{Proof. }
Since $U$ and $U'$ have the same sign of energy we can assume they are HW
representations of $\tilde{S}_n\simeq \widetilde{SO}_0(2,n-1)$.
Let $\lambda$ and $\lambda'$ their respective HW. Let $\nu=0$ (resp. 1) if $n$
is even (resp. odd). Then $n+2=2r+\nu$ and the rank $r'$ of
$\mathcal{S}_n^{\mathbb{C}}$ is given by $r'=r-(1-\nu)$, thus
$\mathcal{S}_n^{\mathbb{C}}$ and $\mathcal{C}_n^{\mathbb{C}}$ have the same
rank if and only if $n$ is odd. Let $(e_i)_{1\le i\le r'}$ be the canonical
basis of $\mathbb{C}^{r'}$ and let:
\begin{equation} \label{E: Delta}
   \Delta_{n+1}^+ =
     \begin{cases}
\{ e_i\pm e_j , 1\le i<j\le r' \}     &\text{if $n+1$ is even;} \\
 \{ e_i\pm e_j , 1\le i<j\le r' \} \cup  \{e_j , 1\le j\le r' \}
&\text{if $n+1$ is odd.}  \\
     \end{cases}
\end{equation}
Then $\Delta_{n+1}^{+}$ defines a set of positive roots for
$\mathcal{S}_n^{\mathbb{C}}$.
Thus if $U$ is a HW irreducible representation of $\tilde{S}_n$ with HW
$\lambda=(\lambda_1,\ldots,\lambda_{r'})$ then a weight $\mu$ of $U$ has
the form:
\begin{align} \label{E: mu}
\mu = &\lambda - \sum_{\alpha \in \Delta_{n+1}^+}
\sum_{p_\alpha \in \mathbb{N}}p_\alpha\alpha     \\
= &-\Big[E+\sum_{j=2}^{r'}(q_j+p_j)+(1-\nu) m_1\Big]e_1 +  \notag  \\
&\sum_{i=2}^{r'}\Big[\lambda_i+q_i-p_i-\sum_{j=i+1}^{r'}(q_{ij}+p_{ij}) -
(1-\nu) m_i+\sum_{j=2}^{i-1}(q_{ji}-p_{ji})\Big]e_i  \notag
\end{align}
where $E=-\lambda_1$ and \( (q_j)_j,\; (p_j)_j,\; (m_j)_j\; (q_{ij})_{i<j}\;
and \;(p_{ij})_{i<j}\; \) are families of naturel integers, such that
\( \sum_{j=2}^1(q_{j2}-p_{j2})=\sum_{j=r'+1}^{r'}(q_{ir'}+p_{ir'})=0. \)

  Let $\sigma=(-s-r'+2+\nu/2,s,\ldots,s)=(-s-\frac{n-2}{2},s,\ldots,s)$
where $2s\in\mathbb{N}$ (resp. $s=0$ or $1/2$) if $n$ is even (resp. odd) and
let $\Lambda$ be the set of such $\sigma$'s. Then each $S_n$-massless
representation of $\tilde{S}_n$ has at least one element of $\Lambda$
as a HW. Indeed if $U_0$ is an $S_n$-massless representation of $\tilde{S}_n$
then one has:
\begin{align}
  \mbox{$n$ even}  &\Longrightarrow
     \begin{cases}
U_0 \sim D(s+\frac{n-2}{2},s,\ldots\!,s), 2s\in\mathbb{N}\; \text{and}\; s\ne0\\
 \text{or}  \\
  U_0 \sim D(\frac{n-2}{2},0,\ldots\!,0) \oplus D(\frac{n}{2},0,\ldots\!,0),
     \end{cases}
                                  \\
                   &  \notag            \\
  \mbox{$n$ odd}   &\Longrightarrow
     \begin{cases}
  U_0 \sim D(\frac{n-1}{2},\frac{1}{2},\ldots\!,\frac{1}{2}) \oplus
D(\frac{n-1}{2},\frac{1}{2},\ldots\!,\frac{1}{2},-\frac{1}{2})  \\
 \text{or}  \\
U_0 \sim D(\frac{n-2}{2},0,\ldots\!,0) \oplus D(\frac{n}{2},0,\ldots\!,0).
     \end{cases}
\end{align}
Now assume that there exists, for an $S_n$-massless representation $U_0$,
for which $\sigma\in\Lambda$ is a HW, two DS $U$ and $U'$ with HW $\lambda$
and $\lambda'$ respectively. Then it is well known that there exists a weight
$\mu$ of $U$ such that $\sigma=\mu+\lambda'$. Since $\mu$ is given by
\eqref{E: mu} one has, for some $s$:
\begin{equation}  \label{E: wgt 1}
  E+E'+\sum_{j=2}^{r'}(q_j+p_j)+(1-\nu)m_1 =s+r'-2-\nu/2
\end{equation}
 and, for each\; $i, 2\le i\le r'$,
\begin{equation}  \label{E: wgt i}
  \lambda_i+\lambda_i'+q_i-p_i-\sum_{j=i+1}^{r'}(q_{ij}+p_{ij})
                     -(1-\nu)m_i+\sum_{j=2}^{i-1}(q_{ji}-p_{ji}) =s.
\end{equation}
Now assume $n\ge5$. Then $r'\ge3$ and \eqref{E: wgt i} becomes, for $i=2$:
  \begin{equation}  \label{E: wgt 2}
    \lambda_2+\lambda_2'+q_2=p_2+\sum_{j=3}^{r'}(q_{2j}+p_{2j})+(1-\nu)m_2+s.
  \end{equation}
Adding $\lambda_2+\lambda_2'-E-E'$ to both sides of \eqref{E: wgt 1} and using
\eqref{E: wgt 2}, one gets:
  \begin{eqnarray} \label{E: wgt 1,2}
  \sum_{j=3}^{r'}(q_j+p_j)+2p_2+(1-\nu)m_1+(1-\nu)m_2+
\sum_{j=3}^{r'}(q_{2j}+p_{2j})= & &  \notag \\
= \lambda_2-E+\lambda_2'-E'+r'-1-\nu/2. & &
  \end{eqnarray}
Thus one has:
  \begin{equation} \label{E: ineq1}
\sum_{j=3}^{r'}(q_j+p_j)+(1-\nu)m_1+p_2\le\lambda_2-E+\lambda_2'-E'+r'-1-\nu/2
  \end{equation}
Now adding $\lambda_3+\lambda_3'+q_{23}-E-E'$ to both sides of \eqref{E: wgt 1}
and using \eqref{E: wgt i} for $i=3$ one finds:
  \begin{eqnarray} \label{E: wgt 1,3}
  q_2+p_2+2p_3+\sum_{j=4}^{r'}(q_j+p_j)+(1-\nu)m_1+(1-\nu)m_3+
\sum_{j=4}^{r'}(q_{3j}+p_{3j})+{} & & \notag \\
 +p_{23} =\lambda_3-E+\lambda_3'-E'+r'-1-\nu/2+q_{23},\quad & &
  \end{eqnarray}
thus
  \begin{equation} \label{E: ineq2}
   q_2\le\lambda_3-E+\lambda_3'-E'+r'-1-\nu/2+q_{23}.
  \end{equation}
  But from \eqref{E: wgt 1,2} one gets:
  \begin{equation} \label{E: ineq3}
    q_{23}\le\lambda_2-E+\lambda_2'-E'+r'-1-\nu/2,
  \end{equation}
so that
  \begin{equation}  \label{E: ineq4}
  q_2\le\lambda_2+\lambda_3-2E+\lambda_2'+\lambda_3'-2E'+2r'-2-\nu.
  \end{equation}
Finally one has, thanks to \eqref{E: ineq1}, \eqref{E: ineq4} and
\eqref{E: wgt 1} \begin{align} \label{E: ineq5}
  s &= \sum_{j=2}^{r'}(q_j+p_j)+(1-\nu)m_1+E+E'-r'+1+\nu/2  \\
    &\le2\lambda_2-2E+\lambda_3+2\lambda_2'-2E'+\lambda_3'+2r'-2-\nu. \notag
  \end{align}
The right hand side being finite for fixed $U$ and $U'$, only $S_n$-massless
representations whose parameter $s$ satisfies \eqref{E: ineq5}, thus a finite
number, may occur in the reduction of $U\otimes U'$.

  Now unitarity of $U$ and $U'$ implies \cite{A83,EHW83}:
  \begin{align}
                           E&\ge\lambda_2+r'-3/2-\nu/2  \\
     \mbox{and: \qquad}   E'&\ge\lambda_2'+r'-3/2-\nu/2,
  \end{align}
but that is not compatible with \eqref{E: wgt 1,2}. Indeed the left hand side of
 \eqref{E: wgt 1,2} is a naturel integer whereas the right hand one satisfies:
  \begin{equation} \label{E: fin ineq}
     \lambda_2-E+\lambda_2'-E'+r'-1-\nu/2\le-r'+2+\nu/2\le-1+\nu/2<0.
  \end{equation}
\vspace{2mm}

%
%  Remark 1
%
\begin{remark}
{\emph  {Unitarity of $U$ or $U'$ is however possible for $n\ge5$. Indeed, the
$S_n$-massless representation $D(\frac{n-2}{2},0,\ldots,0)\oplus
D(\frac{n}{2},0,\ldots,0)$ is contained in the tensor product of the
$C_{n-1}$-massless representation $U=D(\frac{n-3}{2},0,\ldots,0)$, which is
unitary, by the representation $U'=D(\frac{1}{2},0,\ldots,0)\oplus
D(\frac{3}{2},0,\ldots,0)$, which is not unitary. Another example is given by
the $S_n$-massless representation
$D(\frac{n-1}{2},\frac{1}{2},\ldots,\frac{1}{2})$, the unitary
$C_{n-1}$-massless representation
$U=D(\frac{n-2}{2},\frac{1}{2},\ldots,\frac{1}{2})$ and the non unitary
representation $U'=D(\frac{1}{2},0,\ldots,0)$.}}
\end{remark}

  Now let us look to the other values of $n$. As seen above the case $n=4$
is treated in \cite{FF78}, thus we examine only the cases $n=3$ and $n=2$.

\vspace{2mm }

  Let $n=3$. Then the De Sitter algebra $\mathcal{S}_3\simeq\mathfrak{so}(2,2)$
is isomorphic to $\mathfrak{so}(2,1)\oplus\mathfrak{so}(2,1)$.
The $C_3$-massless representations of the conformal algebra
$\mathcal{C}_3\simeq\mathfrak{so}(2,3)$ are the $\Rac=D(1/2,0)$ and the
$\Di=D(1,1/2)$ or, more shortly, $\mbox{D(1/2+s,s)}$, $s$ being 0 or 1/2.
The $S_3$-massless representations of $\mathcal{S}_3$ are thus
$\mbox{\rst{D(1/2+s,s)}{\mathcal{S}_3}}, s=0$ or 1/2. Having in
mind that an irreducible HW representation of $\mathfrak{so}(2,2)$ is
equivalent to a tensor product (which we write $\boxtimes$) of two irreducible
representations of $\mathfrak{so}(2,1)$
one gets:
  \begin{align}
    \rst{D(1/2,0)}{\mathcal{S}_3} &\sim D'(1/2,0)\oplus D'(3/2,0)   \\
   &\sim D(1/4)\boxtimes D(1/4)\oplus D(3/4)\boxtimes D(3/4)  \notag \\
%\eqref\mbox
{\mathrm {and}\qquad\qquad\qquad \qquad}     &   \notag  \\
       &    \notag  \\
    \rst{D(1,1/2)}{\mathcal{S}_3} &\sim D'(1,1/2)\oplus D'(1,-1/2)  \\
 &\sim D(1/4)\boxtimes D(3/4)\oplus D(3/4)\boxtimes D(1/4).  \notag
  \end{align}
Here we have denoted by $D'(E,j)$ (resp. $D(\alpha)$) the irreducible
representation with HW $(-E,j)$ (resp. $(-\alpha)$) of $\mathfrak{so}(2,2)$
(resp. $\mathfrak{so}(2,1)$). Now a large number of UIRs $\pi$ and $\pi'$
of $\tilde{S}_3$ may have the property that $S_3$-massless
representations are contained in $\pi\otimes\pi'$. To reduce that number we
shall suppose that $\pi$ and $\pi'$ are $C_2$-massless, in analogy with the
4-dimensional case where the Dirac
singletons $\Di$ and $\Rac$ are $C_3$-massless. Then, if one assumes that
$\pi$ and $\pi'$ are HW representations, each one has the form
$D'(\alpha,\pm\alpha)$ where $\alpha>0$. Thus one
must consider the products:
  \begin{align}
D'(\alpha,\pm\alpha)&\otimes D'(\beta,\pm\beta)  \\
 D'(\alpha,\pm\alpha)&\otimes D'(\beta,\mp\beta),\qquad \alpha,\beta>0  \notag
  \end{align}
or, equivalently, the products:
  \begin{align}
\Big[D(0)\boxtimes D(\alpha)\Big]&\otimes\Big[D(0)\boxtimes D(\beta)\Big]  \\
\Big[D(0)\boxtimes D(\alpha)\Big]&\otimes\Big[D(\beta)\boxtimes D(0)\Big]
\notag \\\Big[D(\alpha)\boxtimes D(0)\Big]&\otimes\Big[D(\beta)
\boxtimes D(0)\Big]  \notag \\
\Big[D(\alpha)\boxtimes D(0)\Big]&\otimes\Big[D(0)\boxtimes D(\beta)\Big].
\notag
  \end{align}
Now using
  \begin{equation} \label{E: tens sl2}
    D(\alpha)\otimes D(\beta)\sim\bigoplus_{l=0}^\infty D(\alpha+\beta+l)
  \end{equation}
one finds
  \begin{align}
 D'(\alpha,\pm\alpha)\otimes D'(\beta,\pm\beta)&=
\bigoplus_{l=0}^\infty D'(\alpha+\beta+l,\pm[\alpha+\beta+l])
\label{E: tens sing1} \\
 D'(\alpha,\pm\alpha)\otimes D'(\beta,\mp\beta)&=D'(\alpha+\beta,
\pm[\alpha-\beta]). \label{E: tens sing2}
  \end{align}
Finally it is easily seen that
  \begin{align}
                            D'(1/4,1/4)\otimes D'(1/4,-1/4)&=D'(1/2,0)  \\
                            D'(3/4,3/4)\otimes D'(3/4,-3/4)&=D'(3/2,0)
  \end{align}
thus
  \begin{eqnarray}
    \Big[D'(1/4,1/4)\oplus D'(3/4,3/4)\Big]\otimes\Big[D'(1/4,-1/4)
\oplus D'(3/4,-3/4)\Big]=& &  \\
      =\Big[D'(1/2,0)\oplus D'(3/2,0)\Big]\oplus\Big[D'(1,1/2)
\oplus D'(1,-1/2)\Big].& &   \notag
  \end{eqnarray}
The right hand side is a sum of $S_3$-massless representations.  Using
\eqref{E: tens sing1} and \eqref{E: tens sing2}, one can see that this is a
unique solution (up to equivalence) of the problem
Singleton$\otimes$Singleton = $\oplus S_3$-massless for unitary Dirac
singletons, which are, here, $D'(1/4,\pm1/4)\oplus D'(3/4,\pm3/4)$.
They are not irreducible, but each component is irreducible on
both $S_3$ and $L_3$.

\vspace{2mm}

  Let $n=2$. Then $\mathcal{C}_2\simeq\mathfrak{so}(2,2),
\mathcal{S}_2\simeq\mathfrak{so}(2,1)$
and $\mathcal{L}_2\simeq\mathfrak{so}(1,1)$.
A (HW) $C_2$-massless representation of $\mathcal{C}_2$
has the form $D'(\alpha,\pm\alpha), \alpha>0$. Thus the $S_2$-massless
representations of $\mathcal{S}_2$ have the form
$\rst{D'(\alpha,\pm\alpha)}{\mathcal{S}_2}\sim D(\alpha)$.
Now $\mathcal{C}_1$-masslessness
on $D(\beta)$ (or irreducibility on the 2-Lorentz group $L_2$) implies
$\beta=0$, so that, instead of the $n=4$ and $n=3$ cases, Dirac singletons are
not compatible with $C_1$-masslessness.
But one has:
  \begin{equation}
D(\alpha/2)\otimes D(\alpha/2)\sim D(\alpha)\oplus
\bigoplus_{l=0}^\infty D(\alpha+1+l).
  \end{equation}
Thus $S_2$-massless representations occur in the tensor product of two
$S_2$-massless ones.

%
%            6   Indecomposability. Gupta-Bleuler triplets
%
\begin{center}
\section{Indecomposability. Gupta-Bleuler triplets}  \label{S: GB}
\end{center}

  Gupta-Bleuler triplets are used to quantize gauge theories, in a way similar
to the quantization of (4-dimensional flat) QED. This kind of quantization is
done on an indefinite metric space which carries indecomposable
representations, as in the Gupta-Bleuler quantization of the electromagnetic
field. Let us see how it works in the case of our massless representations.
If $U_2$ is a massless representation of $G_n$ then it can be obtained as a
component of an indecomposable representation. Indeed one can find UIRs
$U_\varepsilon$, $\varepsilon>0$, and $U_3$ such that
$\lim_{\varepsilon\to0}U_\varepsilon$ is a non trivial extension
$U_2\rightarrow U_3$ (i.e. we have an exact sequence
\mbox{$0\rightarrow\mathcal{H}_3\rightarrow\mathcal{H}
\rightarrow\mathcal{H}_2\rightarrow0$} where $\mathcal{H}_i$ is the carrying
space of $U_i, i=2$ or 3). The elements of $\mathcal{H}_3$, the
\emph{gauge states}, are obtained from those of $\mathcal{H}$ by applying a
constraint similar to the Lorentz condition in QED. The elements of
$\mathcal{H}_2$, the \emph{physical states}, are realized on the quotient
$\mathcal{H}/\mathcal{H}_3$. Now the representation $(U_2+U_3,\mathcal{H})$
has no invariant nondegenerate metric, thus covariant quantization is not
possible. But if one extends the representation $U_3$ by $U_2+U_3$ in a
non trivial way (\mbox{$U_3\rightarrow U_2\rightarrow U_3$}) to a bigger space
endowed with an invariant nondegenerate (but indefinite) Hermitian form then
quantization of the gauge theory under construction becomes possible.

  In the following we construct some examples of Gupta-Bleuler triplets for the
 massless representations when $n\ge3$.

%
%   6.1  Massless representations and indecomposability
%
\subsection{Massless representations and indecomposability}

  Let us recall that the massless representations for
$\mathcal{G}_n=\mathfrak{so}(2,n)$ are the \mbox{$C_n$-massless} and the
$S_{n+1}$-massless ones. Below we write them again, according to the
parity of $n$. In analogy with 4-dimensional physics we call the parameter
$s$ the \emph{spin} of the representation.

\vspace{2mm}
\newpage
\emph{Case 1: $n$ is even}

\vspace{2mm}

  $C_n$-massless representations are:
   \begin{equation}
 D(s+\frac{\displaystyle n-2}{\displaystyle 2},s,\ldots,\pm s),
\quad 2s\in\mathbb{N}
   \end{equation}

  $S_{n+1}$-massless representations are:
   \begin{align}
      D(\;\frac{n}{2}\;,\,\frac{1}{2}\;,\ldots,\frac{1}{2}\;)
       &\oplus D(\;\frac{n}{2}\;,\frac{1}{2},\ldots,\frac{1}{2},-\frac{1}{2})
     &&\mbox{\quad for spin $\frac{1}{2}$}  \\
      D(\frac{n-1}{2},0,\ldots,0) &\oplus  D(\frac{n+1}{2},0,\ldots,0)
                                      &&\mbox{\quad for spin 0.}
   \end{align}

\vspace{2mm}

\emph{Case 2: $n$ is odd}

\vspace{2mm}

  $C_n$-massless representations are:
   \begin{equation}
                  D(s+\frac{n-2}{2},s,\ldots,s) \quad s\in\{0,\frac{1}{2}\}
   \end{equation}

  $S_n$-massless representations are:
   \begin{align}
      &D(\frac{n-1}{2},0,\ldots,0) \oplus D(\frac{n+1}{2},0,\ldots,0)
\quad \mbox{for spin 0} \\
   &D(s+\frac{n-1}{2},s,\ldots,s), \quad 2s\in\mathbb{N}\quad
\text{and}\quad s\ge\frac{1}{2}.
   \end{align}

  Some of the  above irreducible representations correspond to the limit of
unitarity \cite{A83,EHW83}. It is the case of the $C_n$-massless ones and,
when $n$ is odd, of the $S_{n+1}$-massless representations for  which $s\ge1$.
Then one can look for indecomposability and Gupta-Bleuler (GB) triplets. That
is what we do in the next subsections (for these representations).

  In the next two subsections the cases of the representations
$D(\frac{n-2}{2},0,\ldots,0)$ and \\ $D(\frac{1}{2}+\frac{n-2}{2},\frac{1}{2},
\ldots,\frac{1}{2})$ are treated without separating
the $n$ even and $n$ odd cases, since those representations are
$C_n$-massless for both $n$ even and $n$ odd. Finally, when $s\ge1$ 
the $C_n$-massless $D(s+\frac{n-2}{2},s,\ldots,s)$
for $n$ even and the $S_{n+1}$-massless $D(s+\frac{n-1}{2},s,\ldots,s)$
for $n$ odd are investigated successively.

%
%    6.2  C_n-masslessness, spin 0
%
\subsection{$C_n$-masslessness, spin 0}

%
%    6.2.1   Reduction of $D(E_0,0,\ldots,0)$ to $\mathfrak{k}_n$ and its
%            indecomposability
%
\subsubsection{Reduction of $D(E_0,0,\ldots,0)$ to $\mathfrak{k}_n$ and its
indecomposability}

  Recall that $\mathfrak{k}_n\simeq\mathfrak{so}(2)\oplus\mathfrak{so}(n)$
is the maximal compact subalgebra of $\mathcal{G}_n$.
Let $\mathcal{G}_n=\mathfrak{k}_n+\mathfrak{p}_n$ the Cartan decomposition of
$\mathcal{G}_n$. Let $\bigl(X_{jk}\bigr)_{-r\le j,k\le r}$ a basis
\footnote{This basis is more appropriate to the triangular decomposition of
$\mathcal{G}_n$ than the $\bigl(M_{ab}\bigr)_{a,b}$ basis.} of
$\mathcal{G}_n^\mathbb{C}$ such that:
  \begin{align}
       X_{jk}\; &= -X_{kj}  \\
{\mathrm {and} \qquad}  [X_{jk},X_{j'k'}] &=
\delta_{j,-j'}X_{kk'}+\delta_{k,-k'}X_{jj'}-
 \delta_{j,-k'}X_{kj'}-\delta_{k,-j'}X_{jk'}
  \end{align}
and let:
  \begin{align}
 \mathfrak{n}^\pm &= \langle\; X_{\pm j,\pm k},  (1-\nu)\le j,k \le r \;\rangle
  +\langle\; X_{\pm j,\mp k},  1\le j<k \le r \;\rangle, \\
  \mathfrak{h}\;\; &= \langle\; X_{-j,j}=H_j,  1\le j \le r \;\rangle.
  \end{align}
Then $\mathcal{G}_n^\mathbb{C}=\mathfrak{n}^++\mathfrak{h}+\mathfrak{n}^-$ is
a triangular decomposition of $\mathcal{G}_n^\mathbb{C}$.

  Let $\mathfrak{p}^\pm=\mathfrak{p}_n^\mathbb{C}\cap\mathfrak{n}^\pm$. Then
$\mathcal{G}_n^\mathbb{C}=\mathfrak{p}^++\mathfrak{k}_n^\mathbb{C}+
\mathfrak{p}^-$. The basis $(X_{jk})_{jk}$ is chosen such that:
  \begin{align}
  \mathfrak{p}^\pm &= \langle\; X_{\pm1,j}, -r\le j\le r \quad\text{and}
\quad |j|\ne1 \;\rangle \\
\mathfrak{k}_n^\mathbb{C}\, &= \langle\; X_{jk}, -r\le j,k\le r
\quad\text{and}   \quad |j|,|k|\ne1 \;\rangle
  \end{align}
The root system $\Delta_{n+2}$ is defined by the set of positive roots
$\Delta_{n+2}^+$ which is given by \eqref{E: Delta}, but with $n+2$
(resp. $r=\crl\frac{n+2}{2}\crr$) instead of $n+1$
(resp. $r'=\crl\frac{n+1}{2}\crr$). The new basis is also chosen such that
in the decomposition
$\mathfrak{n}^\pm=\sum_{\alpha>0}\mathcal{G}_n^{(\pm\alpha)}$ the subspace
$\mathcal{G}_n^{(e_j\pm e_k)}$ is, for $1\le j<k\le r$, generated by
$X_{j,\pm k}$ and, if $n$ is odd, $\mathcal{G}_n^{(e_j)}$ is,
for $1\le j\le r$, generated by $X_{0j}$. The roots which correspond to
$\mathfrak{k}_n^\mathbb{C}$ are the \emph{compact} roots and the others the
\emph{noncompact} ones. The set of positive compact (resp. noncompact) roots
is denoted by $\Delta_{n+2}^{+c}$ (resp. $\Delta_{n+2}^{+n}$).

Let $\lambda=(\lambda_1,\ldots,\lambda_r)$ a $\Delta_{n+2}^{+c}$-dominant
integer weight and let $K(\lambda)$ denote the irreducible (finite dimensional)
HW $\mathfrak{k}_n$-module. We write $N(\lambda)$ for the induced HW
$\mathcal{G}_n$-module, with HW $\lambda$, and $L(\lambda)$ for the irreducible
quotient. The HW vectors for both $N(\lambda)$ and $L(\lambda)$ are,
for simplicity, indentified and denoted by $v_\lambda$.

%
%  Proposition 5
%
\begin{proposition} \label{P: Scal Indec}
Let \( E_0>0, \;\;\lambda=(-E_0,0,\ldots,0),\;\; u_\lambda=D(E_0,0,\ldots,0),
\;\;  \mathbf{Z}~=\sum_{|h|\ne1}X_{-1,h}X_{-1,-h} \in
\mathcal{U}(\mathcal{G}_n^\mathbb{C}) \) and, for   \(l,k\in \mathbb{N},
\;\; v_{lk}=(X_{-1,2})^l\mathbf{Z}^kv_\lambda \in N(\lambda)\). Then
    \begin{equation} \label{E: N}
N(\lambda)=\bigoplus_{l,k=0}^\infty \mathcal{U}(\mathfrak{k}_n^\mathbb{C})v_{lk}
    \end{equation}
  and
    \begin{equation} \label{E: N irr}
         N(\lambda)\text{ is irreducible } \Longleftrightarrow E_0 \notin
\bigl\{ \frac{n}{2}-1,\ldots,\frac{n}{2}-\bcrl\frac{n-1}{2}\bcrr \bigr\}.
     \end{equation}
  Moreover if $E_0=\frac{n}{2}-j$ for some
$j\in\{ 1,\ldots,\crl\frac{n-1}{2}\crr\}$ then
    \begin{align} \label{E: N quot}
L(\lambda)=L(-\frac{n}{2}+j,0,\ldots,0) &\simeq
\qt{$N(\lambda)$}{$\bigoplus_{l,k=0}^\infty \mathcal{U}
(\mathfrak{k}_n^\mathbb{C})v_{l,j+k}$} \\
 &\simeq \bigoplus_{l=0}^\infty \bigoplus_{k=0}^{j-1}\mathcal{U}
(\mathfrak{k}_n^\mathbb{C})v_{lk}. \notag
    \end{align}
\end{proposition}

%
%  Corollary 2
%
\begin{corollary} \label{C: Red u}
   Let us write \( \chi(\mu_1)\otimes\pi(\mu_2,\ldots,\mu_r) \) for the
irreducible representation, with HW $\mu$, of $\mathfrak{k}_n$ on $K(\mu)$.
Then
  \begin{align}
\rst{D(E_0,\ldots,0)}{\mathfrak{k}_n} =
&\bigoplus_{l=0}^\infty\bigoplus_{k=0}^\infty
 \chi(-[E_0+l+2k])\otimes\pi(l,0,\ldots,0)  \\
&\text{\;if\;} E_0\notin \frac{n}{2}-
\bigl\{1,\ldots,\bcrl\frac{n-1}{2}\bcrr\bigr\}, \notag \\
\rst{D(E_0,\ldots,0)}{\mathfrak{k}_n} =
 &\bigoplus_{l=0}^\infty\bigoplus_{k=0}^{j-1}
  \chi(-[E_0+l+2k])\otimes\pi(l,0,\ldots,0)  \\
 &\text{\;if\;} E_0 = \frac{n}{2}-j \text{\;for some\;}
 j\in\bigl\{1,\ldots,\bcrl\frac{n-1}{2}\bcrr\bigr\}. \notag
  \end{align}
\end{corollary}

%
%  Remark 2
%
\begin{remark} \label{R: Indec}
 {\emph{ \begin{enumerate}
     \item  The value $j=1$ corresponds to the $C_n$-massless case:
  $$\rst{D(\frac{n-2}{2},0,\ldots,0)}{\mathfrak{k}_n}=\bigoplus_{l=0}^\infty
                          \chi(-[E_0+l])\otimes\pi(l,0,\ldots,0)$$
  which is a particular case of Proposition \ref{P: Red K}.
     \item  Thanks to the preceding results one can see that indecomposability
arises when $E_0$ reaches the value $\frac{n}{2}-j$ (we use the same
notations)~:
  \begin{equation}
    D(E_0,0,\ldots,0)\underset{E_0\to\frac{n}{2}-j}{\longrightarrow}
                       D(\frac{n}{2}-j,0,\ldots,0)+D(\frac{n}{2}+j,0,\ldots,0).
  \end{equation}
 \end{enumerate}}}
\end{remark}

\emph{Proof of the Proposition. }
The $v_{lk}$'s are maximal vectors for $\rst{D(E_0,0,\ldots,0)}
{\mathfrak{k}_n}$; indeed one has $\big[\mathfrak{n}^+
\cap\mathfrak{k}_n^\mathbb{C},X_{-1,2}\big]=0$ and
$\big[\mathfrak{k}_n^\mathbb{C},\mathbf{Z}\big]=0$, thus
$\mathfrak{n}^+\cap\mathfrak{k}_n^\mathbb{C}v_{lk}=0$.
$N(\lambda)$ is generated by the monomials
$\prod_{\overset{j=-r}{|j|\ne1}}^{r}X_{-1,j}^{q_j}v_\lambda$
where $(q_j)_{|j|\ne1}$ is a family of naturel integers and, if $|j|\ne1$:
\[
X_{-1,j}v_{lk}=
    \begin{cases}
\frac{1}{l+1}X_{-2,j}v_{l+1,k}     &\mbox{if $|j|\ne2,$}   \\
          v_{l+1,k}                       &\mbox{if $j=2,$}       \\
\frac{l}{l+1}v_{l-1,k+1}-\frac{1}{(l+1)(l+n)}
\sum_{|h|\ne1,2}X_{-2,-h}X_{-2,h}v_{l+1,k}  &\mbox{if $j=-2,$}
    \end{cases}
\]
where $v_{-1,k}=0$; thus one has
$\mathfrak{p}^-v_{lk}\subset\mathcal{U}(\mathfrak{k}_n^\mathbb{C})v_{l-1,k+1}+
\mathcal{U}(\mathfrak{k}_n^\mathbb{C})v_{l+1,k}$. Since $\big[\mathfrak{p}^-,
\mathfrak{k}_n^\mathbb{C}\big]\subset\mathfrak{p}^-$ one can conclude that
$$N(\lambda)=\mathcal{U}(\mathfrak{p}^-)v_\lambda\subset
\bigoplus_{l,k=0}^\infty\mathcal{U}(\mathfrak{k}_n^\mathbb{C})v_{lk}.$$
Now $\mathfrak{p}^+=\langle\; X_{1j}, -r\le j\le r\;
\text{and}\;|j|\ne1\;\rangle$ and $|j|\ne1$ implies
$$X_{1j}v_{lk}=\delta_{j,-2}l(E_0+2k+l-1)v_{l-1,k}+2k(E_0-
\frac{n}{2}+k)X_{-1,j}v_{l,k-1},$$
with $v_{-1,k}=v_{l,-1}=0$; thus for a maximal vector for which the weight
is strictly less than $\lambda$, necessarily proportional to some $v_{lk}$,
one must have $k(E_0-\frac{n}{2}+k)=0$
and $l=0$, i.e. $l=0, k\ne0$ and $E_0-\frac{n}{2}+k=0$. $E_0$ being strictly
positive one has $1\le k\le\crl\frac{n-1}{2}\crr$.

  Finally let $j\in\{1,\ldots,\crl\frac{n-1}{2}\crr\}$, $E_0=\frac{n}{2}-j$ and
$K_j=\bigoplus_{l=0}^\infty\bigoplus_{k=j}^\infty\mathcal{U}
(\mathfrak{k}_n^\mathbb{C})v_{lk}$. Then the relation
$$\mathfrak{p}_n^\mathbb{C}v_{lk}\subset\mathcal{U}
(\mathfrak{k}_n^\mathbb{C})\langle\;v_{l-1,k}; v_{l+1,k};
v_{l-1,k+1}; v_{l+1,k+1}\;\rangle$$
implies $\mathcal{U}(\mathcal{G}_n^\mathbb{C})K_j\subset K_j$, so that
$$L(-[\frac{n}{2}-j]),0,\ldots,0)=
\qt{$N(-[\frac{n}{2}-j],0,\ldots,0)$}{$K_j$}.$$

%
%    6.2.2  A Gupta-Bleuler triplet for the $C_n$-massless
%           $D(\frac{n-2}{2},0,\ldots,0)$
%
\subsubsection{A Gupta-Bleuler triplet for the $C_n$-massless
$D(\frac{n-2}{2},0,\ldots,0)$}  \label{SSS: Scal GB}

  Using the preceding notations and results one can see that $D(\frac{n-2}{2}+
\varepsilon,0,\ldots,0)$ sends the operator $\mathbf{Z}$ to zero if
$\varepsilon=0$ but it does not if $\varepsilon\ne0$. It is precisely this fact
which gives us the desired indecomposable representations. Indeed, let
$\varepsilon>0$ and $E_0=\frac{n-2}{2}+\varepsilon$. Then $D(E_0,0,\ldots,0)$
is irreducible, but when $\varepsilon\to0$ one obtains, from Remark
\ref{R: Indec} and for $j=1$, an indecomposable representation:
  \begin{equation}
 D(\frac{n-2}{2}+\varepsilon,0,\ldots,0)\underset{\varepsilon\to0}
{\longrightarrow} D(\frac{n-2}{2},0,\ldots,0)+D(\frac{n+2}{2},0,\ldots,0).
  \end{equation}
In order to construct explicitly a Gupta-Bleuler (GB) triplet\cite{Ar85},
let $\rho>0$ and let:
$$H_\rho^{2,n}=\big\{y,\; y=\sum_{a=-1}^ny^ae_a\in\mathbb{R}^{2,n} \text{\;\;
such that\;\;} y^{\;2}=1/\rho\big\}$$
where $y^{\;2}=y^ay_a=y_{\!-1}^{\;2}+y_0^{\;2}+\mathbf{y}^{\;2}$.
The De Sitter space-time is the universal covering of
$H_\rho^{2,n}$. The action of $\mathcal{G}_n$ on
$\mathcal{C}^\infty$-functions defined on $H_\rho^{2,n}$ is well known:
  \begin{equation}
  U_\lambda(M_{ab})=L_{ab}=y_a\partial_b-y_b\partial_a
  \end{equation}
where $\partial_c=\frac{\partial}{\partial y^c}$. Let
$\partial^{\;2}=\partial^a\partial_a$ and $\delta=y^a\partial_a$. Then one has:
  \begin{equation}
U_\lambda(\mathbf{C}_2)=-\frac{1}{2}L_{ab}L^{ab}=
-y^{\;2}\partial^{\;2}+\delta(\delta+n).
  \end{equation}
Now the resolution of the Laplace-Beltrami equation on $H_\rho^{2,n}$ is
standard \cite{RLN66}. One finds that the following solutions form a
Hilbertian basis for $L_\mu^{2}(H_\rho^{2,n})$, with
$\ud\mu(y)=\frac{1}{\rho^{-1}+\mathbf{y}^{\;2}}\ud t\ud^ny$:
\begin{eqnarray}
 \ \!\!\psi_{k\mathbf{lm}}^{E_0}(t,\mathbf{y})\!&=&\!\!\!
\left[\rho^{-(2k+E_0-\frac{n-2}{2})}\frac{\Gamma(k+E_0+l)\Gamma(k+1)}
{\Gamma(k+\frac{n}{2}+l)\Gamma(k+E_0-\frac{n-2}{2})}\right]^{1/2}\!\! \times\\
 &\times& e^{-i(E_0+l+2k)t}\;\;(\rho^{-1} +
\mathbf{y}^{\;2})^{-\frac{E_0+l}{2}}\;(\mathbf{y}^{\;2})^{\frac{l}{2}}\;\;
\times  \notag \\
&\times&\mathrm{P}_k^{(l+\frac{n-2}{2},E_0-\frac{n}{2})}\!\!\!
\left(\frac{\rho^{-1}-\mathbf{y}^{\;2}}{\rho^{-1}+\mathbf{y}^{\;2}}\right)
\mathrm{Y}_\mathbf{m}^\mathbf{l}\!\!\left(\frac{\mathbf{y}}
{\sqrt{(\mathbf{y}^{\;2})}}\right),\notag
\end{eqnarray}
where
 $\mathrm{P}_k^{(\alpha,\beta)}$ are the Jacobi polynomials,
$\mathbf{l}=(l_2,\ldots,l_{\crl\frac{n+1}{2}\crr})$
and $\mathbf{m}=(m_1,\ldots,m_{\crl\frac{n}{2}\crr})$ are vectors, in
$\mathbb{N}^{r-1+\nu}$ and $\mathbb{N}^{r-1}$ respectively, subject to certain
conditions, $l=l_2$, $\mathrm{Y}_\mathbf{m}^\mathbf{l}$ are the spherical
harmonics on $S^{n-1}$ and
$e^{it}=\Bigl(\frac{y^{-1}+iy^0}{y^{-1}+iy^0}\Bigr)^{1/2}$. The scalar product
we use to normalize these functions is given by:
\begin{equation}
  (\psi,\psi')=\int_{\mathbb{R}^n}\overline{\psi}(y)
         \overleftrightarrow{\:i\partial_t\:}\psi'(y)\frac{\ud^ny}{\rho^{-1}+
                 \mathbf{y}^{\;2}},
\end{equation}
where $\overline{\psi}(y)\overleftrightarrow{A}\psi'(y)=
\overline{A\psi}(y)\psi'(y)+\overline{\psi}(y)A\psi'(y)$.
We extend the functions $\psi_{k\mathbf{l}\mathbf{m}}^{E_0}$ to
$H_+^{2,n}=\cup_{\rho>0}H_\rho^{2,n}$
by fixing the degree of homogeneity: $\delta\psi=-E_0\psi$. Then,
$\psi$ being in the kernel of $\partial^{\;2}$, one has:
\begin{equation}
U_\lambda(\mathbf{C}_2)\psi=E_0(E_0-n)\psi
\end{equation}

  Let:
\[
        x_{\pm j}=
  \begin{cases}
 \vspace{3mm}
 \frac{i}{\sqrt2}(y^{-1}\pm iy^0)      &\mbox{\qquad\qquad if $j=1$,}    \\
 \vspace{3mm}
\frac{1}{\sqrt2}(y^{2j-1}\pm iy^{2j}) &\mbox{\qquad\qquad if $2\le j\le r$,} \\

y^n &\mbox{\qquad\qquad if $n$ is odd and $j=0$,}
  \end{cases}
\]
\[
        \partial_j=\frac{\partial}{\partial x_{-j}}  \text{\qquad and \qquad}
                   \partial_{-j}=\overline{\partial_j}.
\]
Then one has $y^{\;2}=-\sum_{j=-r}^{r}x_{-j}x_j,\; \partial^{\;2}=
-\sum_{j=-r}^{r}\partial_{-j}\partial_j, \;
\delta=\sum_{j=-r}^{r}x_{-j}\partial_j$ and one can choose
\footnote{We use the notations of the preceding subsubsection}
$X_{jk}$ such that:
\begin{equation}
 U_\lambda(X_{jk})=x_k\partial_j-x_j\partial_k.
\end{equation}
Let $\varphi_{_2}(y)=x_1^{-E_0}$. Then $\varphi_{_2}$ is, up to a
multiplicative constant, the maximal vector of $U_\lambda$ and
$\psi_{k\mathbf{l}\mathbf{m}}^{E_0}\in\mathcal{U}(\mathcal{G}_n)\varphi_{_2}$.
Moreover one finds that:
\begin{equation}
(\mathbf{Z}\varphi_{_2})(y)=
-E_0(E_0+1)y^{\;2}x_1^{-E_0-2}-2\varepsilon E_0x_{-1}x_1^{-E_0-1},
\end{equation}
thus
\begin{equation}
  \lim_{\varepsilon\to0}(\mathbf{Z}\varphi_{_2})(y)=
-\frac{n-2}{2}\frac{n}{2}y^{\;2}x_1^{-\frac{n+2}{2}}.
\end{equation}

Now assume $\varepsilon=0$ and let $\varphi_{_1}(y)=x_{-1}x_1^{-\frac{n}{2}}$
and $\varphi_{_3}(y)=y^{\;2}x_1^{-\frac{n+2}{2}}$. Then
\begin{equation}
   \varphi_{_1}\overset{\frac{1}{n}\overline{\mathbf{Z}}}{\mymap}
   \varphi_{_2}\overset{-\frac{2}{n}\frac{2}{n-2}\mathbf{Z}}{\mymap}
   \varphi_{_3}
\end{equation}
where $\overline{\mathbf{Z}}=\sum_{|j|\ne1}X_{1j}X_{1,-j}$ and one has:
\[
 \partial^{\;2}\varphi_{_2}=\partial^{\;2}\varphi_{_3}=0
\]
whereas
\[
\partial^{\;2}\varphi_{_1}=\frac{1}{y^{\;2}}\varphi_{_3}\ne0,
\text{\quad but\quad}  (\partial^{\;2})^{2}\varphi_{_1}=0.
\]
Let $cl(V)$ denotes the closure of any topological space $V$ and let
$\EuScript{H}_i^{(0)}=
cl(\mathcal{U}(\mathcal{G}_n^\mathbb{C})\varphi_{_i}),\\ i$ taking the value
 1, 2 or 3. Then it is not difficult to prove the following.

%
%  Proposition 6
%
\begin{proposition}
 \begin{enumerate}
 \item $\EuScript{H}_1^{(0)}\supset\EuScript{H}_2^{(0)}
  \supset\EuScript{H}_3^{(0)}$ and $\EuScript{H}_i^{(0)},
       i=2$ or $3$, is a closed invariant subspace of $\EuScript{H}_{i-1}$;
 \item  \qt{$\EuScript{H}_1^{(0)}$\!\!}{$\EuScript{H}_2^{(0)}$} and
  $\EuScript{H}_3^{(0)}$ carry the IR $D(\frac{n+2}{2},0,\ldots,0)$, while
  \qt{$\EuScript{H}_2^{(0)}$\!\!}{$\EuScript{H}_3^{(0)}$}
       carries the $C_n$-massless representation $D(\frac{n-2}{2},0,\ldots,0)$.
 \item       \begin{align}
   &\Bigl[U_\lambda(\mathbf{C}_2)+\frac{(n-2)(n+2)}{4}\Bigr]\varphi_{_i}=0
   \text{\quad if \;\; $i=2$ or $3$},   \\
   &\Bigl[U_\lambda(\mathbf{C}_2)+\frac{(n-2)(n+2)}{4}\Bigr]\varphi_{_1}=
   n\varphi_{_3}\ne0, \notag \\
   &\Bigl[U_\lambda(\mathbf{C}_2)+\frac{(n-2)(n+2)}{4}\Bigr]^2\varphi_{_1}=0.
   \notag
       \end{align}
 \item $\lim_{y^{\;2}\to0}\varphi(y)=0 \;\forall\varphi\in\EuScript{H}_3^{(0)}$.
   Thus the $C_n$-massless $D(\frac{n-2}{2},0,\ldots,0)$ may be realized
  irreducibly on the cone $Q^{2,n}=\{y,\; y\in\mathbb{R}^{2,n}\text{\quad
  \emph{such that}\quad}y^{\;2}=0\}$.
 \end{enumerate}
\end{proposition}

%
%  Definition 4
%
\begin{definition}
In analogy with QED on $4$-dimensional Minkowski space we call the elements of
$\EuScript{H}_S^{(0)}=$\qt{$\EuScript{H}_1^{(0)}$\!\!}{$\EuScript{H}_2^{(0)}$}
(resp. $\EuScript{H}_P^{(0)}=$\qt{$\EuScript{H}_2^{(0)}$\!\!}
{$\EuScript{H}_3^{(0)}$}, resp. $\EuScript{H}_G^{(0)}=\EuScript{H}_3^{(0)}$)
 \emph{scalar} (resp. \emph{physical}, resp. \emph{gauge}) \emph{states}.
\end{definition}

%
%  Remark 3
%
\begin{remark} {\emph{
Let $\EuScript{K}^{(0)}$ the closure of the $\mathcal{G}_n^\mathbb{C}$-module
generated by $y\longmapsto x_1^{-\frac{n+2}{2}}$; it carries the IR
$D(\frac{n+2}{2},0,\ldots,0)$. Let $\partial^{\;4}=(\partial^{\;2})^2$ and
let us identify $y^{\;2}$ to the corresponding operator. Then the GB triplet
$$D(\frac{n+2}{2},0,\ldots,0)\longrightarrow D(\frac{n-2}{2},0,\ldots,0)
\longrightarrow D(\frac{n+2}{2},0,\ldots,0)$$
defined by $\varphi_{_1}, \varphi_{_2}$ and $\varphi_{_3}$ may be defined by:
  \begin{align}
     \EuScript{H}_1^{(0)} &=\bigl\{ \text{\emph{positive energy solutions} $f$
\emph{of}\quad} \partial^{\;4}f=0\text{\quad \emph{and}\quad}\delta f=
-\frac{n-2}{2}f \bigr\} \notag \\
     \EuScript{H}_2^{(0)} &=\bigl\{ f\in\EuScript{H}_1^{(0)}
 \text{\quad such that\quad}\partial^{\;2}f=0 \bigr\}     \notag \\
\EuScript{H}_3^{(0)} &=\bigl\{ f\in\EuScript{H}_2^{(0)} \text{\quad such that
\quad} f\in y^{\;2}\EuScript{K}^{(0)}\bigr\}. \notag
  \end{align}
  Now, for $\varphi$ and $\varphi'$ in $\EuScript{H}_1^{(0)}$, define
$(\varphi,\varphi')_1=\int_{S^1\times\mathbb{R}^n}
\overline{\varphi}(y)\overleftrightarrow{\:y^2\partial^2\:}\varphi'(y)
\frac{\ud t\ud^ny}{\rho^{-1}+\mathbf{y}^{\;2}}$ and $(\varphi,\varphi')_2=
\int_{S^1\times S^{n-1}}(\mathbf{y}^{\;2})^{\frac{n-2}{2}}\overline{\varphi}(y)
\varphi'(y)\ud t\ud\Omega$, where $y$ belongs to some $H_\rho^{2,n}$
(resp. $Q^{2,n}$) in the first (resp. second) integral. Then it is not
difficult to choose the constant $c$ such that the form defined by
$\langle\varphi,\varphi'\rangle=(\varphi,\varphi')_1+c(\varphi,\varphi')_2$
is an invariant non degenerate indefinite metric such that
$\langle\varphi_i,\varphi_j\rangle\ne0$ if and only if
$(i,j)\in\{(1,3),(3,1),(2,2)\}$. }}
\end{remark}

%
%  Definition 5
%
\begin{definition}
Again in analogy with $4$-dimensional Minkowskian QED, the condition
$\partial^{\;2}f=0$, on $f\in\EuScript{H}_1^{(0)}$, which fixes the space
$\EuScript{H}_2^{(0)}$ will be called \emph{Lorentz condition};
the equation $\partial^4f=0$ will be called the \emph{dipole equation}.
\end{definition}

%
%    6.3  C_n-masslessness, spin 1/2
%
\subsection{$C_n$-masslessness, spin 1/2}

%
%    6.3.1  Reduction on $\mathfrak{k}_n$ and indecomposability of
%           $D(E_0,\frac{1}{2},\ldots,\frac{1}{2})$
%
\subsubsection{Reduction on $\mathfrak{k}_n$ and indecomposability of
$D(E_0,\frac{1}{2},\ldots,\frac{1}{2})$}
  The following result is known; see for example\cite{A83,EHW83}.

%
%  Proposition 7
%
\begin{proposition}
 $D(E_0,\frac{1}{2},\ldots,\frac{1}{2})$ is unitarizable if and only if
$E_0\ge\frac{n-1}{2}$.
\end{proposition}
Here we consider only the unitary case, i.e. $E_0\ge\frac{n-1}{2}$.

%
%  Proposition 8
%
\begin{proposition} \label{P: Red,Indec,1/2}
  Let $\lambda=(-E_0,\frac{1}{2},\ldots,\frac{1}{2})$ and recall that
$\nu=0$ (resp.$1$) if $n$ is even (resp. odd).
  \begin{enumerate}
\item  If $E_0>\frac{n-1}{2}$ then $D(E_0,\frac{1}{2},\ldots,\frac{1}{2})$
is irreducible and one has:
      \begin{align}
 \rst{D(E_0,\frac{1}{2},\ldots,\frac{1}{2})}{\mathfrak{k}_n}=
&\bigoplus_{l,k=0}^\infty\chi\bigl(-[E_0+l+2k]\bigr)\otimes\pi
\bigl(\frac{1}{2}+l,\frac{1}{2},\ldots,\frac{1}{2}\bigr) \oplus \\
\oplus &\bigoplus_{l,k=0}^\infty\chi \bigl(-[E_0+l+2k+1]\bigr)\otimes
\pi\bigl(\frac{1}{2}+l,\ldots,\frac{1}{2},\nu-\frac{1}{2}\bigr). \notag
       \end{align}
    \item  If $E_0=\frac{n-1}{2}$ then $N(\lambda)$ is not simple;
it contains a maximal submodule isomorphic to
\mbox{$L(-\frac{n+1}{2},\frac{1}{2},\ldots,\frac{1}{2},\nu-\frac{1}{2})$}
which carries the UIR
$D(\frac{n+1}{2},\frac{1}{2},\ldots,\frac{1}{2},\nu-\frac{1}{2})$.
The irreducible one $D(\frac{n-1}{2},\frac{1}{2},\ldots,\frac{1}{2})$
is carried by the quotient.
  \end{enumerate}
\end{proposition}
\emph{Proof. }
  \begin{enumerate}
\item If $E_0>\frac{n-1}{2}$ then $D(E_0+\frac{1}{2},0,\ldots,0)\otimes
D(-\frac{1}{2},\frac{1}{2},\ldots,\frac{1}{2})=D(E_0,\frac{1}{2},
\ldots,\frac{1}{2})\oplus \\ D(E_0+1,\frac{1}{2},\ldots,\frac{1}{2},
\nu-\frac{1}{2})$.
If we denote by $v_\sigma$ the maximal vector of
$D(-\frac{1}{2},\frac{1}{2},\ldots,\frac{1}{2})$ one finds that, for
$l,k\in\mathbb{N}$, the vectors $v_{lk}\otimes v_\sigma$ and
$v_{lk}\otimes(X_{-1,[\nu-1]r}v_\sigma)$ generate a submodule
(of the tensor product) isomorphic to $L(\lambda)$.

   \item Now assume $E_0=\frac{n-1}{2}$ and let $\mathbf{Y}^\nu=
\frac{1}{\nu+1}X_{-1,[\nu-1]r}-\sum_{j=2}^rX_{-1,j}X_{-j,[\nu-1]r}$.
Then one can see that $\mathbf{Y}^\nu(v_{00}\otimes v_\sigma)$
generates an irreducible submodule of
$\mathcal{U}(\mathcal{G}_n^\mathbb{C})(v_{00}\otimes v_\sigma)$
isomorphic to $L(-\frac{n+1}{2},\frac{1}{2},\ldots,\nu-\frac{1}{2})$ while
$D(\frac{n-1}{2},\frac{1}{2},\ldots,\frac{1}{2})$ is carried by the quotient
\qt{$\mathcal{U}(\mathcal{G}_n^\mathbb{C})(v_{00}\otimes v_\sigma)$}
{$\mathcal{U}(\mathcal{G}_n^\mathbb{C})\mathbf{Y}^\nu(v_{00}\otimes v_\sigma)$}.
  \end{enumerate}

%
%    6.3.2  A Gupta-Bleuler triplet for $D(\frac{n-1}{2},\frac{1}{2},
%           \ldots,\frac{1}{2})$
%
\subsubsection{A Gupta-Bleuler triplet for $D(\frac{n-1}{2},\frac{1}{2},
               \ldots,\frac{1}{2})$}
  Let $\varepsilon\ge0$ such that $E_0=\frac{n-1}{2}+\varepsilon$.
Proposition \ref{P: Red,Indec,1/2} says that if $\varepsilon=0$ then
$\mathbf{Y}^\nu$ is sent to 0 by
$U_\lambda=D(E_0,\frac{1}{2},\ldots,\frac{1}{2})$.
Now assume $\varepsilon>0$, then $U_\lambda$ is irreducible; but when
$\varepsilon\to0$ one obtains an indecomposable representation:
\begin{equation}
     D\bigl(\frac{n-1}{2}+\varepsilon,\frac{1}{2},\ldots,\frac{1}{2}\bigr)
  \underset{\varepsilon\to0}{\longrightarrow}
     D\bigl(\frac{n-1}{2},\frac{1}{2},\ldots,\frac{1}{2}\bigr)+
  D\bigl(\frac{n+1}{2},\frac{1}{2},\ldots,\frac{1}{2},\nu-\frac{1}{2}\bigr).
\end{equation}
To construct a Gupta-Bleuler triplet we need explicit realizations of the
representations concerned. Let $\sigma=(\frac{1}{2},\ldots,\frac{1}{2})$ and
let, if $n$ is even, $\sigma^-=(\frac{1}{2},\ldots,\frac{1}{2},-\frac{1}{2})$.
We denote by $S_\sigma$ the irreducible spinor representation
$D(-\frac{1}{2},\frac{1}{2},\ldots,\frac{1}{2})$ and, when $n$ is even,
by $S_{\sigma^-}$ the irreducible one
$D(-\frac{1}{2},\frac{1}{2},\ldots,\frac{1}{2},-\frac{1}{2})$.
Let $\EuScript{S}_+$ be the carrier space of $S_\sigma$ and $\EuScript{S}_-$
the carrier one of $S_{\sigma^-}$ when $n$ is even (resp. \{0\} when $n$ is
odd). Finally let $\EuScript{S}=\EuScript{S}_+\oplus\EuScript{S}_-$ be the
spinor module of $\mathcal{G}_n$.

  Let $\gm_{-1},\ldots,\gm_{2r-2}$ be $2r$ matrices in
$\mathfrak{gl}(\EuScript{S})$ such that $[\gm_a,\gm_b]_+=2\eta_{ab}$
\footnote{We identify the identity of
$\mathfrak{gl}(\EuScript{S})$ with 1.}, where $[A,B]_+=AB+BA$, and let
$\gm_{2r-1}\in\mathbb{C}\gm_{-1}\cdots\gm_{2r-2}$ such that
$\gm_{2r-1}^{\;2}=-1$.\mbox{ Then:}
\[
 [\gm_a,\gm_b]_+=2\eta_{ab}\qquad\forall a,b\in\{-1,\ldots,n\}.
\]
The following realization of $S_\sigma$ on $\EuScript{S}$ is well known:
\[
M_{ab}\mymap S_{ab}=\frac{1}{4}[\gm_a,\gm_b]=\frac{1}{2}(\gm_a\gm_b-\eta_{ab}).
\]
Later we shall need also the generators $\om_j$ defined by:
\[
       \om_{\pm j}=
  \begin{cases}
  \vspace{3mm}
 \frac{i}{\sqrt2}(\gm_{-1}\pm i\gm_0)    &\mbox{\qquad\qquad if $j=1$,}  \\
  \vspace{3mm}
 \frac{1}{\sqrt2}(\gm_{2j-1}\pm i\gm_{2j})
&\mbox{\qquad\qquad if $2\le j\le r$,}  \\

\gm_n     &\mbox{\qquad\qquad if $n$ is odd and $j=0$,}
  \end{cases}
\]
Thus one has:
\[
 [\om_j,\om_k]_+=-2\delta_{j,-k}\qquad\forall j,k\in\{-r,\ldots,r\}
\]
and the preceding realization of $S_\sigma$ may be written:
\[
 X_{jk}\mymap\frac{1}{4}[\om_j,\om_k]=\frac{1}{2}(\om_j\om_k+\delta_{j,-k}).
\]

  We realize $D(E_0,\frac{1}{2},\ldots,\frac{1}{2})$ on spinor fields
 $\Psi:H_+^{2,n}\longrightarrow\EuScript{S}_+$ such that
\[
\partial^{\;2}\Psi=0 \text{\qquad and\qquad} \delta\Psi=-(E_0+\frac{1}{2})\Psi.
\]
The action of $\mathcal{G}_n$ on spinor fields is given by
$U_\lambda(M_{ab})=L_{ab}+S_{ab}$. Let:
\[
  \emph{$\ys$}=\sum_{a=-1}^ny^a\gm_a=\sum_{j=-r}^rx_{-j}\om_j
\text{\quad and\quad}
   \emph{$\ds$}=\sum_{a=-1}^n\partial^a\gm_a=-\sum_{j=-r}^r\partial_{-j}\om_j.
\]
Then\footnote{We identify $y^{\;2}$ with the function $y\longmapsto y^{\;2},\,
\emph{$\ys$}$ with $y\longmapsto \emph{$\ys$}$, and so on.}:
\begin{align}
U_\lambda(\mathbf{C}_2)\Psi&=\Bigl[-y^{\;2}\partial^{\;2}+\delta(\delta+n+1)+
\frac{(n+1)(n+2)}{8}-\emph{$\ys\ds$}\Bigr]\Psi \\
&=\Bigl[(E_0+\frac{1}{2})(E_0-\frac{1}{2}-n)+
\frac{(n+1)(n+2)}{8}-\emph{$\ys\ds$}\Bigr]\Psi.\notag
\end{align}
It is easy to prove the following Lemma.

%
%  Lemma 1
%
\begin{lemma} \label{L: prop y, d}
\begin{enumerate}
 \item  $\ys$ and $\ds$ commute with the action of $\mathcal{G}_n$;
 \item  $[\ys,\ds]_+=2\delta+n+2$;
 \item  $-y^{\;2}\partial^{\;2}=\ys\,\ds ( \ys\,\ds-2\delta-n)$;
 \item  if $\varepsilon>0$ then $(-2\varepsilon)^{-1}(\ds\ys-2)$ and
$(-2\varepsilon)^{-1}\ys\ds$ are projectors on the irreducible subspaces of
the tensor product $L(-[E_0+\frac{1}{2}],0,\ldots,0)\otimes L(\frac{1}{2},
\dots,\frac{1}{2})$, namely the spaces $L(-E_0,\frac{1}{2},\ldots,\frac{1}{2})$
and $L(-[E_0+1],\frac{1}{2},\ldots,\frac{1}{2},\nu-\frac{1}{2})$ respectively.
\end{enumerate}
\end{lemma}

  Let us consider the spinor fields $\Psi_2$ and $\Psi_3$ defined by
$$
 \Psi_2(y)=x_1^{-E_0-\frac{1}{2}}v_\sigma \text{\qquad and\qquad}
 \Psi_3(y)=\emph{$\ys$} x_1^{-E_0-\frac{3}{2}}\om_{[\nu-1]r}v_\sigma.
$$
Then one has:
$$
 L(-[E_0+\frac{1}{2}],0,\ldots,0)\otimes L(\frac{1}{2},\ldots,\frac{1}{2})
\simeq  \mathcal{U}(\mathcal{G}_n^\mathbb{C})\Psi_2\oplus\mathcal{U}
(\mathcal{G}_n^\mathbb{C})\Psi_3.
$$
Moreover let $\Psi_1(y)=x_1^{-E_0-\frac{1}{2}}\om_{-1}\om_{[\nu-1]r}$. Then:
\begin{equation}
    \mathbf{Y}^\nu\Psi_2=\frac{1}{2}(E_0+\frac{1}{2})\Psi_3 -
\varepsilon\frac{1}{2}\Psi_1,
\end{equation}
thus
\begin{equation}
\lim_{\varepsilon\to0}\mathbf{Y}^\nu\Psi_2=\frac{1}{2}(E_0+\frac{1}{2})\Psi_3.
\end{equation}

 {} From now on we assume $\varepsilon=0$, i.e. $E_0=\frac{n-1}{2}$. Then:
\begin{equation}
       \Psi_1\overset{-\frac{1}{2-\nu}X_{1,-[\nu-1]r}}{\mymap}
       \Psi_2\overset{\frac{4}{n}\mathbf{Y}^\nu}{\mymap}\Psi_3.
\end{equation}
Let $\EuScript{H}_i^{(1/2)}=cl(\mathcal{U}(\mathcal{G}_n^\mathbb{C})\Psi_i),
\, i$ being 1,\, 2 or 3.
The next proposition is not difficult to prove.

%
%  Proposition 9
%
\begin{proposition}
  \begin{enumerate}
     \item  $\EuScript{H}_1^{(1/2)}\supset\EuScript{H}_2^{(1/2)}
\supset\EuScript{H}_3^{(1/2)}$ and $\EuScript{H}_i^{(1/2)}, i=2$ or $3$,
is a closed invariant subspace of $\EuScript{H}_{i-1}^{(1/2)}$;
 \item  \qt{$\EuScript{H}_1^{(1/2)}$\!\!}{$\EuScript{H}_2^{(1/2)}$} \,and
\,$\EuScript{H}_3^{(1/2)}$ carry\, the\, IR\, \mbox{$D(\frac{n+1}{2},
\frac{1}{2},\ldots,\nu-\frac{1}{2})$}, while\\
\qt{$\EuScript{H}_2^{(1/2)}$\!\!}{$\EuScript{H}_3^{(1/2)}$} carries the
$C_n$-massless representation $D(\frac{n-1}{2},\frac{1}{2},\ldots,
\frac{1}{2})$;
     \item \begin{align}
\ds\Psi_i &=0   \text{\quad if\quad $i=2$ or $i=3$}, \\
\ys\ds\Psi_1 &=n\Psi_3\ne0 \text{\quad but\quad} (\ys\ds)^2\Psi_1=0.\notag
           \end{align}
\item  $\lim_{y^{\;2}\to0}(\ys\Psi)(y)=0\; \forall \Psi\in
\EuScript{H}_3^{(1/2)}$ and\, $\lim_{y^{\;2}\to0}(\ys\Psi_2)(y)\ne0$.
Thus the $C_n$-massless representation $D(\frac{n-1}{2},\frac{1}{2},\ldots,
\frac{1}{2})$ may be realized irreducibly on the cone $Q^{2,n}$.
  \end{enumerate}
\end{proposition}

%
%  Definition 6
%
\begin{definition}
The elements of the space\, $\EuScript{H}_S^{(1/2)}=
\qt{$\EuScript{H}_1^{(1/2)}$\!\!}{$\EuScript{H}_2^{(1/2)}$}$
(resp. $\EuScript{H}_P^{(1/2)}= \\ \qt{$\EuScript{H}_2^{(1/2)}$\!\!}
{$\EuScript{H}_3^{(1/2)}$}$, resp. $\EuScript{H}_G^{(1/2)}=
\EuScript{H}_3^{(1/2)}$) are called \emph{scalar} (resp. \emph{physical},
resp. \emph{gauge}) \emph{states}.
\end{definition}

%
%  Remark 4
%
\begin{remark} {\emph{
  Let ${\EuScript{K}}^{(1/2)}$ be the closure of the
$\mathcal{G}_n^\mathbb{C}$-module generated by the field
$y\longmapsto \Phi(y)= \\ x_1^{-\frac{n+1}{2}}\om_{[\nu-1]r}v_\sigma$; it
carries the IR
$D(\frac{n+1}{2},\frac{1}{2},\ldots,\frac{1}{2},-\frac{1}{2})$.
Then the Gupta-Bleuler triplet
$$
  D\bigl(\frac{n+1}{2},\frac{1}{2},\ldots,\frac{1}{2},
\nu-\frac{1}{2}\bigr)\longrightarrow  D\bigl(\frac{n-1}{2},\frac{1}{2},
\ldots,\frac{1}{2}\bigr)\longrightarrow  D\bigl(\frac{n+1}{2},\frac{1}{2},
\ldots,\frac{1}{2},\nu-\frac{1}{2}\bigr)
$$
 defined by $\Psi_1, \Psi_2$ and $\Psi_3$ may be redefined by:
  \begin{align}
\EuScript{H}_1^{(1/2)} &=\bigl\{\text{positive energy solutions of}\;\,
\partial^{\;2}\Psi=0, \;\delta\Psi=-\frac{n}{2}\Psi\;\, \text{and}\;\,
(\ys\ds)^2\Psi=0\bigr\}, \notag \\
\EuScript{H}_2^{(1/2)} &=\bigl\{\Psi\in\EuScript{H}_1^{(1/2)}
\text{\quad such that\quad}\ds\Psi=0\bigr\}, \notag \\
\EuScript{H}_3^{(1/2)} &=\bigl\{\Psi\in\EuScript{H}_2^{(1/2)}
\text{\quad such that\quad}\Psi\in\ys\EuScript{K}^{(1/2)}\bigr\}.
  \end{align}
Now, for $\Psi$ and $\Psi'$ in $\EuScript{H}_1^{(1/2)}$, define
$(\Psi,\Psi')_1=\rho^{-1}\int_{S^1\times\mathbb{R}^n}\Psi^*(y)
\overleftrightarrow{\:\ys\ds\:}\Psi'(y)\frac{\ud t\ud^ny}{\rho^{-1}+
\mathbf{y}^{\;2}}$ and $(\Psi,\Psi')_2=\int_{S^1\times S^{n-1}}
(\mathbf{y}^{\;2})^{\frac{n}{2}}\Psi^*(y)\Psi'(y)\ud t\ud\Omega$, $y$
being in some $H_\rho^{2,n}$ (resp. $Q^{2,n}$) in the first
(resp. second) integral. Again it is not difficult to choose
the constant $c$ such that the form defined by $\langle\Psi,\Psi'\rangle=
(\Psi,\Psi')_1+c(\Psi,\Psi')_2$ is an invariant non degenerate indefinite
metric such that $\langle\Psi_i,\Psi_j\rangle\ne0$
if and only if $(i,j)\in\{(1,3),(3,1),(2,2)\}$.}}
\end{remark}

%
%  Definition 7
%
\begin{definition}
  The equation $\ds\Psi=0$, which fixes the space $\EuScript{H}_2^{(1/2)}$,
will be called the \emph{Lorentz condition}.
\end{definition}

%
%    6.4  Indecomposability and GB triplets for spin s\ge1
%
\subsection{Indecomposability and GB triplets for  spin $s\ge1$ }
  We assume in this subsection that $s\ge1$ and $2s\in\mathbb{N}$.

%
%    6.4.1  Indecomposability of $D(E_0,s,\ldots,s,s_\nu)$
%
\subsubsection{Indecomposability of $D(E_0,s,\ldots,s,s_\nu)$}

  Let $\lambda=(-E_0,s,\ldots,s,s_\nu)$, where $|s_\nu|=s$ and, if $n$ is odd,
$s_\nu\ge0$.

%
%  Proposition 10
%
\begin{proposition}
  \begin{enumerate}
    \item $D(E_0,s,\ldots,s,s_\nu)$ is unitarizable
$\Longleftrightarrow E_0\ge\frac{n-2+\nu}{2}+s$;
    \item if $E_0>\frac{n-2+\nu}{2}+s$ then $N(\lambda)$ is simple;
    \item if $E_0=\frac{n-2+\nu}{2}+s$ then $N(\lambda)$ contains, up to a
multiplicative constant, a unique maximal vector of weight
$(-E_0-1,s,\ldots,s,s_\nu-\frac{s_\nu}{s})$; it is given by
\;$\mathbf{Y}_{-1,-\frac{s_\nu}{s}r}^\nu v_\lambda$,\; where
   \begin{align}
\mathbf{Y}_{-1,\pm r}^0 &=2sX_{-1,\pm r}-\sum_{j=2}^rX_{-1,j}X_{-j,\pm r},
\notag \\
\mathbf{Y}_{-1,-r}^1    &=2sX_{-1,-r}+2X_{-1,0}X_{-r,0}- \notag \\
&-\frac{2(s-1)}{2s-1}\sum_{j=2}^rX_{-1,j}X_{-j,-r}-\frac{2}{2s-1}
\sum_{j=2}^rX_{-1,j}X_{-j,0}X_{-r,0}. \notag
   \end{align}
  \end{enumerate}
\end{proposition}
 Since, for $n$ even, the treatment of $U_{\lambda}$ is similar for both signs
of $s_\nu$ we shall consider from now on that $s_\nu=s$.

\emph{Proof of the Proposition. }
For the  first two items see \cite{A83,EHW83}. For the last one, a maximal
vector of weight
$(-E_0-1,s,\ldots,s,s-1)$ for $n$ even  has the general form
$$
      v'= \bigl(aX_{-1,-r}+\sum_{j=2}^rb_jX_{-1,j}X_{-j,-r}\bigr)v_\lambda,
$$
and $\mathfrak{n}^+v'=0$ implies $b_j=-\frac{a}{2s}$ for each $j$. The same
technique works for odd $n$.

%
% Remark 5
%
\begin{remark} {\emph{
  The situation for $s\ge1$, for both $n$ even and $n$ odd, is more complicated
than for the spin $0$ and spin $1/2$ cases. Indeed more than one submodule
for $N(\lambda)$ exists when $E_0=\frac{n-2+\nu}{2}+s$, thus it is a priori
possible to construct very different examples of Gupta-Bleuler triplets
$U'\longrightarrow U_\lambda\longrightarrow U'$ with $U'$ unitary.}}
\end{remark}

%
%   6.4.2  A GB triplet for $D(\frac{n-2+\nu}{2}+s+i,s,\ldots,s,s-i),
%          i=1$ or $2$
%
\subsubsection{A GB triplet for $D(\frac{n-2+\nu}{2}+s+i,s,\ldots,s,s-i),
               i=1$ or $2$}
  Let $E_0=\frac{n-2+\nu}{2}+s+\varepsilon, \varepsilon\ge0$.
To realize our Gupta-Bleuler triplet we need explicitely the representations
\mbox{$D(E_0,s,\ldots,s)$} and \mbox{$D(E_0+1,s,\ldots,s,s-1)$}, especially
for $\varepsilon=0$. Both of them are contained in the reduction of the tensor
 product $D(E_0+s,0,\ldots,0)\otimes D(-s,s,\ldots,s)$. The representation
$S_{[2s\sigma]}=D(-s,s,\ldots,s)$ itself is contained in the tensor power
$S_\sigma^{\otimes 2s}$ of the irreducible spinorial representation.

  We define the action of $M_{ab}\in\mathcal{G}_n$ on a tensor
$v_1\otimes\cdots\otimes v_{2s}\in\EuScript{S}_+^{\otimes2s}$ by:
\begin{align} \label{E: act tens}
S_{ab}(v_1\otimes\cdots\otimes v_{2s})&=\sum_{t=1}^{2s}v_1\otimes\cdots\otimes
\frac{1}{4}[\gm_a,\gm_b]v_t\otimes\cdots\otimes v_{2s}  \\
    &=\sum_{t=1}^{2s}S_{ab}^{(t)}(v_1\otimes\cdots
   \otimes v_t\otimes\cdots\otimes v_{2s}). \notag
\end{align}
Let ${\gm_a}^{(t)}$ be defined on the tensors\footnote{Recall that
$\EuScript{S}_-=\{0\}$ for $n$ odd.} of $\EuScript{S}^{\otimes2s}=
(\EuScript{S}_+\oplus\EuScript{S}_-)^{\otimes2s}$  by:
$${\gm_a}^{(t)}(v_1\otimes\cdots\otimes v_{2s})=
v_1\otimes\cdots\otimes\gm_av_t\otimes\cdots\otimes v_{2s}.$$
Then the action defined in \eqref{E: act tens} may be written more simply:
\begin{equation}
 M_{ab}\mymap S_{ab}=\sum_{t=1}^{2s}S_{ab}^{(t)}=
  \sum_{t=1}^{2s}\frac{1}{4}\bigl[{\gm_a}^{(t)},{\gm_b}^{(t)}\bigr].
\end{equation}

  Let Sym$(\EuScript{S}_+^{\otimes2s})$ be the space of symmetric tensors in
$\EuScript{S}_+^{\otimes2s}$ and let $\gm^{(t)}\!\cdot\!\gm^{(t')}=
{\gm^a}^{(t)}{\gm_a}^{(t')}=-\sum_{j=-r}^r\om_{-j}^{(t)}\om_j^{(t')}$.

%
%  Proposition 11
%
\begin{proposition}
  $S_{[2s\sigma]}=D(-s,s,\ldots,s)$ is realized irreducibly on the space:
\[
   V^S=\mathrm{Sym}(\EuScript{S}_+^{\otimes2s})\cap\Bigl[\,
\bigcap_{\underset{t\ne t'}{t,t'}}\ker\bigl(\gm^{(t)}\!\cdot\!
\gm^{(t')}-\nu\bigr)\,\Bigr].
\]
\end{proposition}

  We realize the unitary representations of interest on tensor-spinors
\mbox{ $\Psi:H_+^{2,n}\longrightarrow V^S$} such that
$$
  \partial^{\;2}\Psi=0 \text{\qquad and\qquad} \delta\Psi=-(E_0+s)\Psi.
$$
To this effect, we define the action of $\mathcal{G}_n$ on them by
$M_{ab}\longmapsto L_{ab}+S_{ab}$. Let

$$
   \emph{$\yst{t}$}=y^a{\gm_a}^{(t)}=\sum_{j=-r}^rx_{-j}\om_j^{(t)}
\text{\quad and\quad} \emph{$\dst{t}$}=\partial^a{\gm_a}^{(t)}=
-\sum_{j=-r}^r\partial_{-j}\om_j^{(t)}
$$
then one has
\begin{align}
 U_\lambda(\mathbf{C}_2)\Psi&=\Bigl[-y^{\;2}\partial^{\;2}+\delta(\delta+n+2s)+
rs(s+r-1+\nu)-\sum_{t=1}^{2s}\emph{$\yst{t}\dst{t}$}\Bigr]\Psi \\
 &=\Bigl[(E_0+s)(E_0-s-n)+rs(s+r-1+\nu)-\sum_{t=1}^{2s}
\emph{$\yst{t}\dst{t}$}\Bigr]\Psi.\notag
\end{align}

%
%  Lemma 2
%
\begin{lemma}
  \begin{enumerate}
 \item For fixed $t$,\:$\yst{t}$ and $\dst{t}$ satisfy the three first items
of Lemma\ref{L: prop y, d};
 \item if\quad $t\ne t'$\quad then\quad $[\yst{t},\yst{t'}]=0$\quad
and\quad $[\dst{t},\dst{t'}]=0$;
 \item if\quad $t\ne t'$\quad then\quad $[\dst{t},\yst{t'}]=
{\gm}^{(t)}\cdot{\gm}^{(t')}\;(=\nu \text{\;on\;} V^S)$.
\end{enumerate}
\end{lemma}

  Let us define, for non negative integers $k, l$ and spinors $v_1,\ldots,v_k$,
symmetric tensors in $\EuScript{S}^{\otimes2s}$ by~:
\begin{align}
    v_1\cdots v_k &=\frac{1}{k!}\sum_{\tau\in\mathfrak{S}_k}\tau(v_1)\otimes
\cdots\otimes\tau(v_k), \notag \\
         v_1^l &=\underbrace{v_1\cdots v_1}_{l\:\mathrm{terms}}
\end{align}
and let $\Psi_1, \Psi_2$ and $\Psi_3$ be defined by:
\begin{align}
\Psi_1(y) &=x_1^{-E_0-s}\Bigl[\bigl(\om_{-1}\om_{-r}v_\sigma\bigr)
v_\sigma-\nu\bigl(\om_{-1}v_\sigma\bigr)\bigl(\om_{-r}v_\sigma\bigr)
\Bigr]v_\sigma^{2s-2}, \notag \\
\Psi_2(y) &=x_1^{-E_0-s}v_\sigma^{2s}, \notag \\
\Psi_3(y) &=x_1^{-E_0-s-1}\Bigl[\bigl(\emph{$\ys$}\om_{-r}v_\sigma\bigr)
v_\sigma-\nu\bigl(\emph{$\ys$}v_\sigma\bigr)\bigl(\om_{-r}v_\sigma\bigr)\Bigr]
v_\sigma^{2s-2}. \notag
\end{align}
Then one has
$$
\mathcal{U}(\mathcal{G}_n^\mathbb{C})\Psi_2\oplus\mathcal{U}
(\mathcal{G}_n^\mathbb{C})\Psi_3\subset
L(-[E_0+s],0,\ldots,0)\otimes L(s,\ldots,s)
$$
and one finds that:
\begin{equation}
     \mathbf{Y}_{-1,-r}^\nu\Psi_2=s(E_0+s)\Psi_3-\varepsilon s\Psi_1,
\end{equation}
thus
\begin{equation}
       \lim_{\varepsilon\to0}\mathbf{Y}_{-1,-r}^\nu\Psi_2=s(E_0+s)\Psi_3.
\end{equation}

 {} From now on we assume $E_0=\frac{n-2+\nu}{2}+s$. Then:
\begin{equation}
 \Psi_1\overset{-\frac{1}{2}X_{1,r}}{\mymap}
 \Psi_2\overset{\frac{2}{s(n-2+\nu+4s)}\mathbf{Y}_{-1,-r}^\nu}{\mymap}\Psi_3.
\end{equation}
Let $\EuScript{H}_i^{(s)}=cl(\mathcal{U}(\mathcal{G}_n^\mathbb{C})\Psi_i)$,
$i$ being equal to 1, 2 or 3. The next proposition is straightforward:

%
%  Proposition 12
%
\begin{proposition}
  \begin{enumerate}
    \item  $\EuScript{H}_1^{(s)}\supset \EuScript{H}_2^{(s)}\supset
\EuScript{H}_3^{(s)}$ and
$\EuScript{H}_i^{(s)}, i=2$ or $3$, is a closed invariant subspace of
$\EuScript{H}_{i-1}^{(s)}$.
 \item  \qt{$\EuScript{H}_1^{(s)}$\!\!}{$\EuScript{H}_2^{(s)}$} \,and \,
$\EuScript{H}_3^{(s)}$ carry\, the\, IR\, \mbox{$D(\frac{n+\nu}{2}+s,s,
\ldots,s,s-1)$}, while\\ \qt{$\EuScript{H}_2^{(s)}$\!\!}
{$\EuScript{H}_3^{(s)}$} carries the representation
$D(\frac{n-2+\nu}{2}+s,s,\ldots,s)$;
    \item \begin{align}
\dst{t}\Psi_i&=0 \;\forall t\in \{1,\ldots,2s\} \text{\quad if\quad $i=2$ or
$i=3$}\text{;} \\
\sum_{t=1}^{2s}\yst{t}\dst{t}\Psi_1&=
(n-2+\nu+4s)\Psi_3\ne0 \text{\quad but\quad}\bigl(\sum_{t=1}^{2s}\yst{t}
\dst{t}\bigr)^2\Psi_1=0 \text{;}     \notag
  \end{align}
 \item  $\lim_{y^{\;2}\to0}(\yst{1}\cdots\yst{2s}\Psi)(y)=0\;
\forall \Psi\in\EuScript{H}_3^{(s)}$ and\,
 $\lim_{y^{\;2}\to0}(\yst{1}\cdots\yst{2s}\Psi_2)(y)\ne0$. Thus the
representation $D(\frac{n-2+\nu}{2}+s,s,\ldots,s)$ may be realized irreducibly
 on the cone $Q^{2,n}$.
  \end{enumerate}
\end{proposition}

%
%  Definition 8
%
\begin{definition}
   The elements of the space\, $\EuScript{H}_S^{(s)}=
\qt{$\EuScript{H}_1^{(s)}$\!\!}{$\EuScript{H}_2^{(s)}$}$
(resp. $\EuScript{H}_P^{(s)}=\qt{$\EuScript{H}_2^{(s)}$\!\!}
{$\EuScript{H}_3^{(s)}$}$, resp. $\EuScript{H}_G^{(s)}=\EuScript{H}_3^{(s)}$)
are called \emph{scalar} (resp. \emph{physical}, resp. \emph{gauge})
 \emph{states}.
\end{definition}

  Let, for $t\in\mathbb{N}, v^{t}\otimes(v\wedge v')=v^{t+1}
\otimes v'-v^{t}\otimes v'\otimes v$, and let
$\bigl(\tau_{(t,t')}\bigr)_{t\le t'}$ be the system of generators (permutations
$t\leftrightarrow t'$ if $t\ne t'$ and identity if $t=t'$) of the
group-algebra of $\mathfrak{S}_{2s}$. Let
$$
\EuScript{Y}=
  \begin{cases} \vspace{4mm}
 \frac{1}{2s}\Bigl[\sum_{1\le t\le2s}\tau_{(t,2s)}\Bigr]\emph{$\yst{2s}$}
&\text{\quad if $n$ is even};\\
 \frac{1}{2s(2s-1)}\Bigl[\sum_{1\le t\le2s-1}\tau_{(t,2s-1)}+& \\
\qquad\quad +\sum_{1\le t<t'\le2s-1}\tau_{(t,2s)}\tau_{(t',2s-1)}\Bigr]
\bigl[\emph{$\yst{2s-1}$}-\emph{$\yst{2s}$}\bigr] &\text{\quad if $n$ is odd}.
  \end{cases}
$$
Finally let
$$
\Phi(y)= x_1^{-\frac{n+\nu}{2}-2s}\times
     \begin{cases}
 v_\sigma^{2s-1}\otimes\om_{-r}v_\sigma   &\text{\qquad if $n$ is even}; \\
 v_\sigma^{2s-2}\otimes(v_\sigma\wedge\om_{-r}v_\sigma) &\text
{\qquad if $n$ is odd}.
     \end{cases}
$$
As in the cases $s=0$ and $s=1/2$ we have here:

%
%  Remark 6
%
\begin{remark} {\emph{
  Let ${\EuScript{K}}^{(s)}$ be the closure of the simple
$\mathcal{G}_n^\mathbb{C}$-module generated by the field $\Phi$; it carries
the IR $D(\frac{n+\nu}{2}+s,s,\ldots,s,s-1)$. Then the Gupta-Bleuler triplet \\
 \mbox{$D\bigl(\frac{n+\nu}{2}+s,s,\ldots,s,s-1\bigr)\!\rightarrow\!
  D\bigl(\frac{n-2+\nu}{2}+s,s,\ldots,s\bigr)\!\rightarrow\!
  D\bigl(\frac{n+\nu}{2}+s,s,\ldots,s,s-1\bigr)$}\\
defined by $\Psi_1, \Psi_2$ and $\Psi_3$ may be redefined by:
  \begin{align}
 \EuScript{H}_1^{(s)} =\bigl\{&\;\text{positive energy solutions of}\;\,
\partial^{\;2}\Psi=0,\\ &\delta\Psi=(-\frac{n-2+\nu}{2}-2s)\Psi\;\, \text{and}
\;\, \bigl(\sum_{t=1}^{2s}\yst{t}\dst{t}\bigr)^2\Psi=0\;\bigr\}, \notag \\
\EuScript{H}_2^{(s)} =\bigl\{&\;\Psi\in\EuScript{H}_1^{(s)}\text{\quad
such that\quad}\dst{t}\Psi=0\quad\forall t\in\{1,\ldots,2s\}\;\bigr\},\notag \\
\EuScript{H}_3^{(s)} =\bigl\{&\;\Psi\in\EuScript{H}_2^{(s)}\text{\quad
such that\quad}\Psi\in\EuScript{Y}\EuScript{K}^{(s)}\;\bigr\}.\notag
  \end{align}
Now, as in spin $0$ and spin $1/2$ cases, one can find an invariant non
degenerate form on $\EuScript{H}_1^{(s)}$. Let \;$(\Psi,\Psi')_1=
(\rho^{-1})^{2s+\frac{\nu}{2}}\int_{S^1\times\mathbb{R}^n}\Psi^*(y)
\overleftrightarrow{\:\sum_{t=1}^{2s}\yst{t}\dst{t}\:}\Psi'(y)
\frac{\ud t\ud^ny}{\rho^{-1}+\mathbf{y}^{\;2}}$\; and \;$(\Psi,\Psi')_2= \\
\int_{S^1\times S^{n-1}}(\mathbf{y}^{\;2})^{\frac{n-2+\nu}{2}+2s}\Psi^*(y)
\Psi'(y)\ud t\ud\Omega$, $\Psi$ and $\Psi'$ being in $\EuScript{H}_1^{(s)}$
and $y$ belongs to some $H_\rho^{2,n}$ (resp. $Q^{2,n}$) in the first (resp.
second) integral. Again it is not difficult to choose the constant $c$ such that
the form defined by $\langle\Psi,\Psi'\rangle=(\Psi,\Psi')_1+c(\Psi,\Psi')_2$
is an invariant non degenerate indefinite metric for which
$\langle\Psi_i,\Psi_j\rangle\ne0$
if and only if $(i,j)\in\{(1,3),(3,1),(2,2)\}$.}}
\end{remark}

%
%  Definition 9
%
\begin{definition}
  The system of equations $\dst{t}\Psi=0$, $t\in\{1,\ldots,2s\}$, which fixes
 the space $\EuScript{H}_2^{(s)}$, will be called the \emph{Lorentz condition}.
\end{definition}

%
%    6.5  Further remarks on GB triplets
%
\subsection{Further remarks on GB triplets}

  The above considerations show that the true generalization of Dirac singletons
from 4-dimensional De Sitter space to space-time in dimension $n\ge5$ are in
fact the $C_{n-1}$-massless representations. Indeed, though ($C_{n-1}$-massless)
$\otimes$($C_{n-1}$-massless) does not contain $S_n$-massless representations
in general (this is true only for $n$=3 or 4), their restriction to the
$n$-Lorentz group $\widetilde{SO}_0(1,n-1)$ is irreducible and they are,
together with their conjugates, the only representations with that property.
Thus they contract to UIRs of the $n$-Poincar\'e group which are trivial on
the translations and their weight diagram is very degenerate, in some sense
1-dimensional.

\vspace{2mm}

  Let us look at indecomposability and construction of gauge theories with GB
quantization. The most interesting case is when $n\ge5$. GB triplets are
easily constructed for $C_{n-1}$-massless representations and for arbitrary
spin $s$ ($2s\in\mathbb{N}$). But for $S_n$-massless representations, which
represent massless particles on $n$-De Sitter space-time, the situation is
different. Indeed, if $n$ is even, $S_n$-massless representations exist for
arbitrary spin $s$, but one can construct a GB triplet, with our method, only
for $s\ge1$, because for $s=$ 0 or 1/2 no indecomposability arises around the
corresponding highest (or lowest) weight. Nevertheless these representations
occur (once) in the tensor product of a $C_{n-1}$-massless representation by a
non unitary one, namely $D(\frac{n-3}{2},0,\ldots,0)\otimes
\bigl[D(\frac{1}{2},0,\ldots,0)\oplus D(\frac{3}{2},0,\ldots,0)\bigr]$ for spin
0 and $D(\frac{n-2}{2},\frac{1}{2},\ldots,\frac{1}{2})\otimes
D(\frac{1}{2},0,\ldots,0)$ for spin 1/2, for which it seems that construction of
GB triplets is possible. Thus one can hope to construct, for even $n\ge6$, a
gauge theory analogous to that of $D(1,0)\oplus D(2,0)$ and
$D(\frac{3}{2},\frac{1}{2})$ with the usual Dirac singletons when $n=4$.

\vspace{2mm}

  Now assume $n$ is odd. Then $S_n$-massless representations exist only for
spin 0 or 1/2 and, again, one cannot construct a GB triplet, but they are
contained in the reduction of the same tensor products as in the even case
(here the last term in the unitary factor is $\pm\frac{1}{2}$).
However, unlike the latter, the representations $D(\frac{1}{2},0,\ldots,0)$
and $D(\frac{3}{2},0,\ldots,0)$, which are also below the unitary limit, cannot
be naturally considered as quotients of extensions, while this is the case
for the $C_{n-1}$-massless representations $D(\frac{n-3}{2},0,\ldots,0)$ and
$D(\frac{n-2}{2},\frac{1}{2},\ldots,\frac{1}{2},\pm\frac{1}{2})$. Thus for odd
$n$ only one factor in the tensor product has naturally GB triplets. One can
ask the question of what would be the analogue of a gauge theory in this
context.
\newpage
%
%                             7  Discussion
%
\begin{center}
\section{Discussion}
\end{center}

  First we recall that in the $n=2$ case the situation is drastically
different from $n\ge3$. Indeed though $C_2$-masslessness is easily defined,
there is no good notion of spin and there exist infinitely many $C_2$-massless
nonequivalent representations with equivalent restrictions to the
2-Poincar\'e group; but this is not the case of the restrictions to the
2-De Sitter group, locally isomorphic to $SO_0(2,1)$.
Note however that in this case (as is well known) the full
conformal group is infinite-dimensional.
\vspace{2mm}

  We have shown here that most properties of massless representations
are, in some sense, independent of the space-time dimension $n$. But if
$n\ge3$, the property of occuring in the tensor product of two UIRs of the
same energy sign is true only for $n=3$ and $n=4$. An interpretation is that
compositeness of massless particles is not possible in De Sitter space-time
with dimension $n\ge5$. Concerning the Gupta-Bleuler quantization, it can
be seen that for general $n\ge3$ the construction of triplets works with
no problem, but for a given massless representation there is no unique
solution to the construction of a Gupta-Bleuler triplet.
\vspace{2mm}

  There is however some ambiguity when one tries to generalize to the
$n$-dimensional Anti-De Sitter space
$\widetilde{SO}_0(2,n-1)/\widetilde{SO}_0(1,n-1)$, $n\ge5$,
the notion of masslessness. For $n\ge5$ there is no canonical definition
of a massless representation of the $n$-De Sitter group
$\tilde{S}_n = \widetilde{SO}_0(2,n)$, especially for ``spin" $s\ge 1$.
Indeed the rank of the compact subalgebra $\mathfrak{so}(n-1)$ of the De
Sitter algebra is $\ge 2$, instead of 1 in the $n=4$ case.
Thus there are several slightly different alternatives to describe
massless particles in De Sitter world in higher dimensions, which coincide for
$n=4$. The two extreme are, for spin $s\ge1$ ($2s\in\mathbb{N}$),
$U(s)=D(s+\frac{n-2}{2},s,\ldots,s,\epsilon s)$, where $|\epsilon|=1$ if
$n$ is odd and $\epsilon=1$ if $n$ is even, and $U'(s)=D(s+n-3,s,0,\ldots,0)$
if $s$ is an integer or $U'(s)=D(s+n-3,s,1/2,\ldots,1/2,\epsilon/2)$ if
$s-1/2$ is an integer. The former are what we call here $S_n$-massless
representations for $n$ even (for $n$ odd, $s\ge1$ there are no $S_n$-massless
representations in our sense). The latter have very recently been called,
when $n=5$, massless (in the bulk) in \cite{FerFr98a,FerFr98b}.

 In what follows we shall compare somewhat in detail various properties of
both alternatives. In order to do this we need first to look more closely at
the notion of masslessness in the $n$-dimensional Minkowski space
$\mathbb{R}^{1,n-1}$. On this basis we then compare the notions of
Anti-De Sitter masslessness in $n\ge5$ dimensions, also in both alternatives.
\vspace{2mm}

  As in the 4-dimensional Minkowski space one needs, for massless
representations of the $n$-Poincar\'e group $\tilde{P}_n$, the mass operator to
be zero (and the representation non trivial). Thus the massless representation,
say $U^P$, of interest must be induced by a UIR of a subgroup which is a
semi-direct product of a subgroup of the $n$-Lorentz group isomorphic to the
Euclidean group $\tilde{E}(n-2)=\widetilde{SO}(n-2)\ltimes\mathbb{R}^{n-2}$
by the group of space-time translations $\mathbb{R}^{1,n-1}$. Moreover, for
physical reasons, it seems reasonnable to eliminate the ``continuous spin'' in
the inducing representation, i.e. we assume that the Euclidean group part of
the inducing representation is trivial on the translations subgroup
$\mathbb{R}^{n-2}$. It is thus finite dimensional and essentially determined
by a UIR $\pi_\lambda$ of $\widetilde{SO}_0(n-2)$ with HW $\lambda$.

A first problem (not appearing in the comparison between our approach and that
of \cite{FerFr98a,FerFr98b}) is that the choice of $\lambda$, to define a
spin $s\ge1$, is not unique for $n\ge6$; to make this choice easier one can
use the physically sensible fact that the wave equations for massless
particles are invariant under the action of the $n$-conformal group
$\tilde{C}_n$. Thus we may add the following \emph{extension condition},
always satisfied for massless representations when $n=4$: $U^P$ extends to a
UIR $\Hat{U}$ of the $n$-conformal group. An interesting consequence of this
(strong) condition is that $\lambda$ depends now on a unique parameter $s$ such
that \cite{AL97}:
\begin{equation} \label{E: lam}
  \lambda=
    \begin{cases}
          (s,\ldots,s,\epsilon s), 2s\in\mathbb{N} \text{ and $|\epsilon|=1$
                                     if $n$ is even}, \\
          (s,\ldots,s), s\in\{0,1/2\} \text{ if $n$ is odd}.
    \end{cases}
\end{equation}
The bad news, with this condition, is that in \emph{odd} dimensional
space-times (and already for $n=5$), one can define naturally neither massless
particles with spin $s\ge1$ nor helicity. Nevertheless we call here,
for uniformity of presentation, a mass zero representation of the
$n$-Poincar\'e group which satisfies the extension condition a
\emph{massless discrete helicity representation} (MDHR).

%
% Remark 7
% 
\begin{remark}\emph{
 In fact one may define helicity for a large class of
 representations  of $\widetilde{P}_n$, especially when $n$ is even. 
%But we think that, in this case and for the representation 
%$\pi_{(s,\ldots,s,\epsilon s)}$, helicity is more naturally defined.
Indeed, take $n$ even and denote by 
$\varepsilon=(\varepsilon_{\mu_1\cdots\mu_n})_{0\le\mu_1,\ldots,\mu_n\le n-1}$ 
the completely skew-symmetric tensor such that
$\varepsilon_{01\cdots(n-1)}=1$. Define the generalized Pauli-Lubanski
vector by:
$$
W_\mu=-\frac{(i/2)^{r-2}(-1)^\frac{(r-2)(r-3)}{2}}{(r-2)!}
    \sum_{\nu_1,\ldots,\nu_{n-1}}\varepsilon_{\mu\nu_1\cdots\nu_{n-1}}
    M^{\nu_1\nu_2}\cdots M^{\nu_{n-3}\nu_{n-2}}P^{\nu_{n-1}}
$$
where the $M^{\nu\nu'}$'s and the $P^\nu$'s stand for the generators of
the $n$-Poincar\'e
algebra and where $r-2$ is the rank of the maximal compact subgroup
 $K_{n-2}$ of $\widetilde{E}(n-2)$ (in fact, we have denoted by $r$ the 
rank of the
Lie algebra of the $n$-conformal group $G_n$). Then one can show 
easily that in a massless (UI) representation of $\widetilde{P}_n$
one has, if $\lambda=(\lambda_3,\ldots,\lambda_r)$ is the HW of the 
irreducible representation $\pi_\lambda$ of $K_{n-2}$:
$$
W_\mu = \epsilon (\lambda_3+r-3)\cdots(\lambda_{r-1}+1)|\lambda_r|P_\mu
$$
where $\epsilon$ is the sign of $\lambda_r$. Thus one may define naturally
helicity thanks to this relation provided that $\lambda_r\ne0$, in which 
case it could not. Let us look at two examples. 
The first one is when all the 
components of $\lambda$ are equal to $s$ modulo $\epsilon$. In this case one 
has, in the same conditions as above:
$$
W_\mu = \epsilon(s+r-3)\cdots(s+1)sP_\mu.
$$
This relation not only fixes the sign of the helicity but determines also
the spin $s$.
The second is when $\lambda_3=s$ and the other components equal to $\sigma$
modulo $\epsilon$ where $\sigma$, being $0$ or $1/2$, is such that 
$s-\sigma$ is an integer. Then one has:
$$
 W_\mu = \epsilon(s+r-3)\cdots(\sigma+1)\sigma P_\mu, 
$$
which equals $0$ when $s$ is an integer. Thus this relation, in this 
example, is not appropriate to define helicity for two kinds of particles
(bosons and fermions) simultaneousely. For these reasons and some others
(for example the conformal invariance of equations) we prefer the first 
example to induce representations which describe massless particles in 
the Minkowski space-time $M_n$.}  
\end{remark}

  If one drops the extension condition, for example for odd $n$, then
$\pi_\lambda$ with $\lambda=(s,\ldots,s,\epsilon s)$ or even $\lambda=
(s,0,\ldots,0)$ or $(s,1/2,\ldots,1/2,\epsilon1/2)$ and $s\ge1$ may be used to
induce a ``massless'' representation $U^P$ in order to represent a massless
particle with spin $s$ in the $n$-dimensional Minkowski space.

  Now consider the following masslessness conditions, analogous to the $n=4$
ones:\\
\indent (a) Massless representations of the $n$-De Sitter group $\tilde{S}_n$ 
contracts
smoothly to a MDHR of the $n$-Poincar\'e group $\tilde{P}_n$;\\
\indent (b) The unique extension to a UIR $\Hat{U}$ of the $n$-conformal group 
of any
MDHR of the $n$-Poincar\'e group is such that $\rst{\Hat{U}}{\tilde{S}_n}$ is
precisely a massless representation of $\tilde{S}_n$;\\
\indent (c) For $s\ge1$ one may construct a gauge theory on the $n$-dimensional
Anti-De Sitter space for massless particles, quantizable by the use of an
indefinite metric and a GB triplet;\\
\indent (d) The massless representations are such that the physical signals
 propagate
on the Anti-De Sitter light cone.

We define also what we shall call here a \emph{singleton property} (SP):
\begin{center}
Singleton $\otimes$ Singleton\emph{\quad contains\quad}Massless representations.
\end{center}
Then it was proved in \cite{AL97} that representations of $\tilde{S}_n$ which
satisfy conditions (a),(b),(c) above have the form $U_0$ with $U_0$ given by
(15) and (16) in Section 5. Thus there is no $S_n$-massless representation
for $s\ge1$ and $n$ odd. This, of course, is related to what happens in flat
$n$-dimensional space where (for $n$ odd) the spin can be only 0 or 1/2
(for a MDHR). As a consequence, when $s\ge1$, conditions (a),(b),(c) are
relevant for $U(s)$ only if $n$ is even and not at all for $U'(s)$ ($n\ge5$).
We conjecture that a representation of the $n$-de Sitter group that satisfies
(a),(b),(c) must satisfy also condition (d) (this will be proved in a
forthcoming paper).

  Unfortunately (for $n\ge5$) property (SP) is not satisfied by $S_n$-massless
representations, i.e. by representations which satisfy (a),(b),(c), as shown
in Proposition 4.

  Now if one drops one (or more) of the masslessness conditions then one can
define other representations of $\tilde{S}_n$ to be massless ones, i.e. to
represent massless particles on the $n$-dimensional Anti-de Sitter space-time.
Between (a), (b) and (c) (and probably (d)) only dropping the stronger condition
(b) has actually an effect because (b) implies both (a) and (c) (and probably
(d)). Indeed if one drops (b) then things change radically. For example let $D=
D(E_0,\lambda_2,\ldots,\lambda_{r'})$ a UIR of $\tilde{S}_n$. If the weight
$(-E_0,\lambda_2,\ldots,\lambda_{r'})$ reaches the limit of unitarity then
usually one obtains an indecomposable representation from which one may
construct a GB triplet and then a gauge theory. Thus condition (c) is still
satisfied by a large number of representations. A contraction of $D$ to a
UIR $U^P$ of $\tilde{P}_n$ is usually possible but the contracted
representation $U^P$ is not a MDHR in general. Let us look at the example
of $U'(s)$, $s\ge1$. From what precedes $U'(s)$ does not satisfy (b).
Moreover $U'(s)$ contracts naturally to a representation $U_\lambda^P$,
of the $n$-Poincar\'e group, for which $\pi_\lambda=\pi_{(s,0,\ldots,0)}$
or $\pi_\lambda=\pi_{(s,1/2,\ldots,1/2,\epsilon/2)}$, depending on whether
$s$ or $s-1/2$ is integer. But $U^P$ is not a MDHR. Even more, $U'(s)$
does not, in general, contract to a MDHR, for which the weight
$\lambda$ of $\pi_\lambda$ must satisfy the relation \eqref{E: lam},
especially when $n$ is odd, case of which the allowed spin is 0 or 1/2.

 Among the 3 masslessness conditions we have studied so far, only (c) is
totally satisfied by $U'(s)$. Indeed let $\sigma=0$ (resp. 1/2) if
$s$ (resp. $s-1/2$) is integer ($\ge 1$). Then the representation
$D(s+n-3+\varepsilon,s,\sigma,\ldots,\sigma,\epsilon\sigma)$
becomes indecomposable if $\varepsilon\to0$ (see some examples in
\cite{FerFr98b} for low values of $n$) and one may construct a GB triplet
{\small
$$D(s+n-2,s-1,\sigma,\ldots,\sigma,\epsilon\sigma) \to
   D(s+n-3,s,\sigma,\ldots,\sigma,\epsilon\sigma) \to
      D(s+n-2,s-1,\sigma,\ldots,\sigma,\epsilon\sigma).
$$}
Moreover, if we consider the ($C_{n-1}$-massless) representations
$D(\frac{n-3}{2},0,\ldots,0)$ and\\$D(\frac{n-2}{2},1/2,\ldots,1/2,\epsilon/2)$
as \emph{singletons}, because they have properties similar to the singletons
Rac and Di (see sections 1, 3 and 4 and subsection 6.5), though they are
\emph{not Dirac singletons} in the sense of Definition 3, then
$U'(s)$ satisfies property (SP) because the following  is true:
\begin{align}
D(\frac{n-3}{2},0,\ldots,0)\otimes &D(\frac{n-3}{2},0,\ldots,0)
                \mathrm{\quad contains } \notag \\
        &\;\;\,\bigoplus_{s=0}^\infty D(n-3+s,s,0,\ldots,0)
                                                      \notag \\
D(\frac{n-3}{2},0,\ldots,0)\otimes &D(\frac{n-2}{2},1/2,\ldots,1/2,\epsilon/2)
               \mathrm{\quad contains } \notag \\
     &\bigoplus_{s-1/2=0}^\infty D(n-3+s,s,1/2,\ldots,1/2,\epsilon/2).
                   \notag
\end{align}

\vspace{5mm}

\textbf{ACKNOWLEDGEMENTS}
\vspace{2mm}

  The author thanks Professors M. Flato and E. Angelopoulos for stimulating
discussions and Professor D.~Sternheimer for helpful criticism of the
manuscript.

\newpage

%
% References
%

\end{document}